\documentclass[ijds,fleqn,nonblindrev]{informs4}

\usepackage[labelfont=bf, labelsep=period, justification=centering]{caption}
\usepackage{float}

\usepackage[T1]{fontenc}
\usepackage[utf8]{inputenc}

\usepackage{mathtools}
\usepackage{amsfonts}
\usepackage{amssymb}
\usepackage{bm}
\usepackage{booktabs}
\usepackage{graphicx}

\usepackage{multirow}

\usepackage[ruled,vlined]{algorithm2e}

\usepackage{tikz}
\usepackage{collcell}

\usepackage{todonotes}

\usepackage{hyperref}
\hypersetup{breaklinks,hyperindex,citecolor=blue,linkcolor=blue}
\usepackage{cleveref}
\usepackage[super]{nth}

\OneAndAHalfSpacedXI

\global\long\def\gg{\,\|\,}%
\global\long\def\R{\mathbb{R}}%

\global\long\def\mba{\bm{a}}%
\global\long\def\mbA{\bm{A}}%
\global\long\def\mbB{\bm{B}}%
\global\long\def\mbC{\bm{C}}%
\global\long\def\mbD{\bm{D}}%
\global\long\def\mbE{\bm{E}}%
\global\long\def\mbI{\bm{I}}%
\global\long\def\mbL{\bm{L}}%
\global\long\def\mbM{\bm{M}}%
\global\long\def\mbP{\bm{P}}%
\global\long\def\mbQ{\bm{Q}}%
\global\long\def\mbU{\bm{U}}%
\global\long\def\mbV{\bm{V}}%
\global\long\def\mbX{\bm{X}}%
\global\long\def\mbZ{\bm{Z}}%
\global\long\def\mbzero{\bm{0}}%
\usepackage{natbib}
 \bibpunct[, ]{(}{)}{,}{a}{}{,}%
 \def\bibfont{\small}%
\TheoremsNumberedThrough     
\ECRepeatTheorems

\EquationsNumberedThrough    

\MANUSCRIPTNO{IJDS-0001-1922.65}

\begin{document}



\RUNTITLE{Low-Rank Robust Subspace Tensor Clustering}

\TITLE{Low-Rank Robust Subspace Tensor Clustering for Metro Passenger Flow Modeling}

\ARTICLEAUTHORS{%
\AUTHOR{Nurretin Dorukhan Sergin}
\AFF{School of Computing and Augmented Intelligence, Arizona State University, Tempe, AZ 85281} 
\AUTHOR{Jiuyun Hu}
\AFF{School of Computing and Augmented Intelligence, Arizona State University, Tempe, AZ 85281}

\AUTHOR{Ziyue Li}
\AFF{Cologne Institute for Information Systems, University of Cologne, Cologne, Germany}
\AFF{EWI gGmbH, University of Cologne, Cologne, Germany, zlibn@wiso.uni-koeln.de}
\AUTHOR{Chen Zhang}
\AFF{Department of Industrial Engineering, Tsinghua University, Beijing, China, \EMAIL{zhangchen01@tsinghua.edu.cn}}

\AUTHOR{Fugee Tsung}
\AFF{Department of Industrial Engineering and Decision Analytics, Hong Kong University of Science and Technology, Hong Kong}
\AFF{Information Hub, Hong Kong University of Science and Technology (Guangzhou), China, \EMAIL{season@ust.hk}}

\AUTHOR{Hao Yan}
\AFF{School of Computing and Augmented Intelligence, Arizona State University, Tempe, AZ 85281, \EMAIL{haoyan@asu.edu}}

} 

\ABSTRACT{%
    Tensor clustering has become an important topic, specifically in spatiotemporal modeling, due to its ability to cluster spatial modes (e.g., stations or road segments) and temporal modes (e.g., time of the day or day of the week). Our motivating example is from subway passenger flow modeling, where similarities between stations are commonly found. However, the challenges lie in the innate high-dimensionality of tensors and also the potential existence of anomalies. 
    This is because the three tasks, i.e.,  dimension reduction, clustering, and anomaly decomposition, are inter-correlated to each other, and treating them in a separate manner will render a suboptimal performance. Thus, in this work, we design a tensor-based subspace clustering and anomaly decomposition technique for \textbf{simultaneously} outlier-robust dimension reduction and clustering for high-dimensional tensors. To achieve this, a novel low-rank robust subspace clustering decomposition model is proposed by combining Tucker decomposition, sparse anomaly decomposition, and subspace clustering. An effective algorithm based on Block Coordinate Descent is proposed to update the parameters. Prudent experiments prove the effectiveness of the proposed framework via the simulation study, with a gain of \textbf{+25\%} clustering accuracy than benchmark methods in a hard case. The interrelations of the three tasks are also analyzed via ablation studies, validating the interrelation assumption. Moreover, a case study in the station clustering based on real passenger flow data is conducted, with quite valuable insights discovered. 
}%


 \HISTORY{This paper was first submitted on Oct 8th, 2022}

\maketitle

%

\section{Introduction} \label{sec:chap3:introduction}

Higher-order tensors have been actively used in research since they have an inclination to successfully preserve the complicated correlation structures of data. A tensor can be defined mathematically as multi-dimensional arrays \citep{kolda2009tensor}. The order of a tensor is the number of dimensions, also known as modes. Tensor clustering is a recent generalization of the basic one-dimensional clustering problem to a high-dimensional version, and it seeks to partition an order-$K$ input tensor into coherent sub-tensors (e.g., slices on one mode) while optimizing some cluster-related measuring criteria \citep{jegelka2009approximation}. 

Tensor clustering has also become an important topic in spatiotemporal modeling \citep{sun2019dynamic, bahadori2014fast, mao2022jointly}, due to its ability to cluster complicated spatial modes (e.g., stations, road segments, etc.) and temporal modes (e.g., time of day or day of week). For example, in the modeling of passenger flow, similarities between spatial elements are commonly found in spatio-temporal data \citep{li2020tensor}. These clusters may be due to natural geographical locations (e.g., neighboring stations) or contextual information (e.g., points of interest) in the context of where the data are collected. For example, in residential areas, stations often have a large number of in-flow passengers in the morning on weekdays; instead, in the business areas, in-flow peaks are expected in the afternoon on weekdays. {\color{black} Exploring metro station clusters can help public transportation management, like operational efficiency, Strategic Planning, Anomaly Detection, Policy Making, and Land Use Planning.} Overall, the following two challenges present for data collected from such complex spatiotemporal systems.

\begin{figure}[t]
    \centering
    \includegraphics[width=0.59\linewidth]{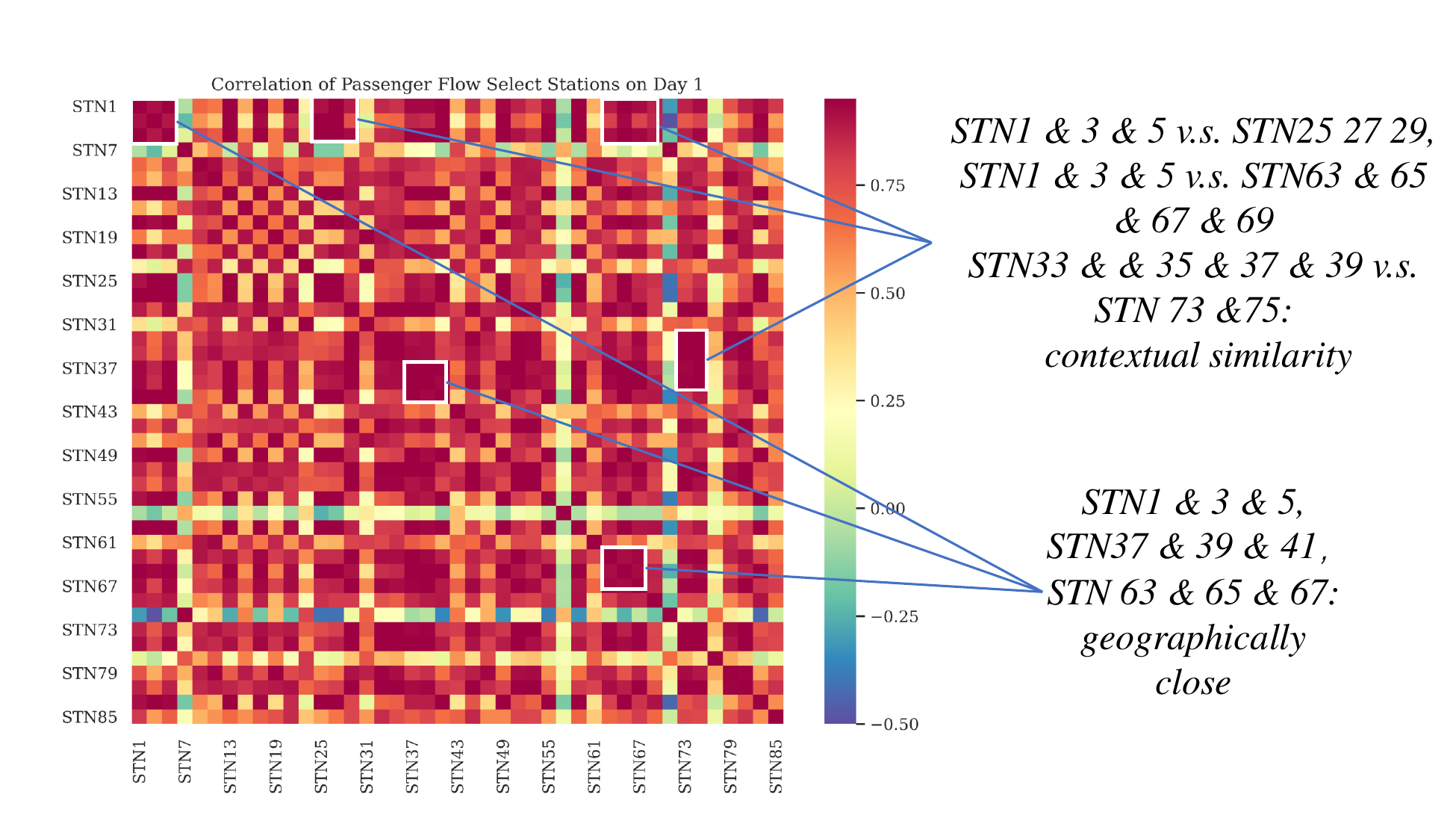}
    \includegraphics[width=0.39\linewidth]{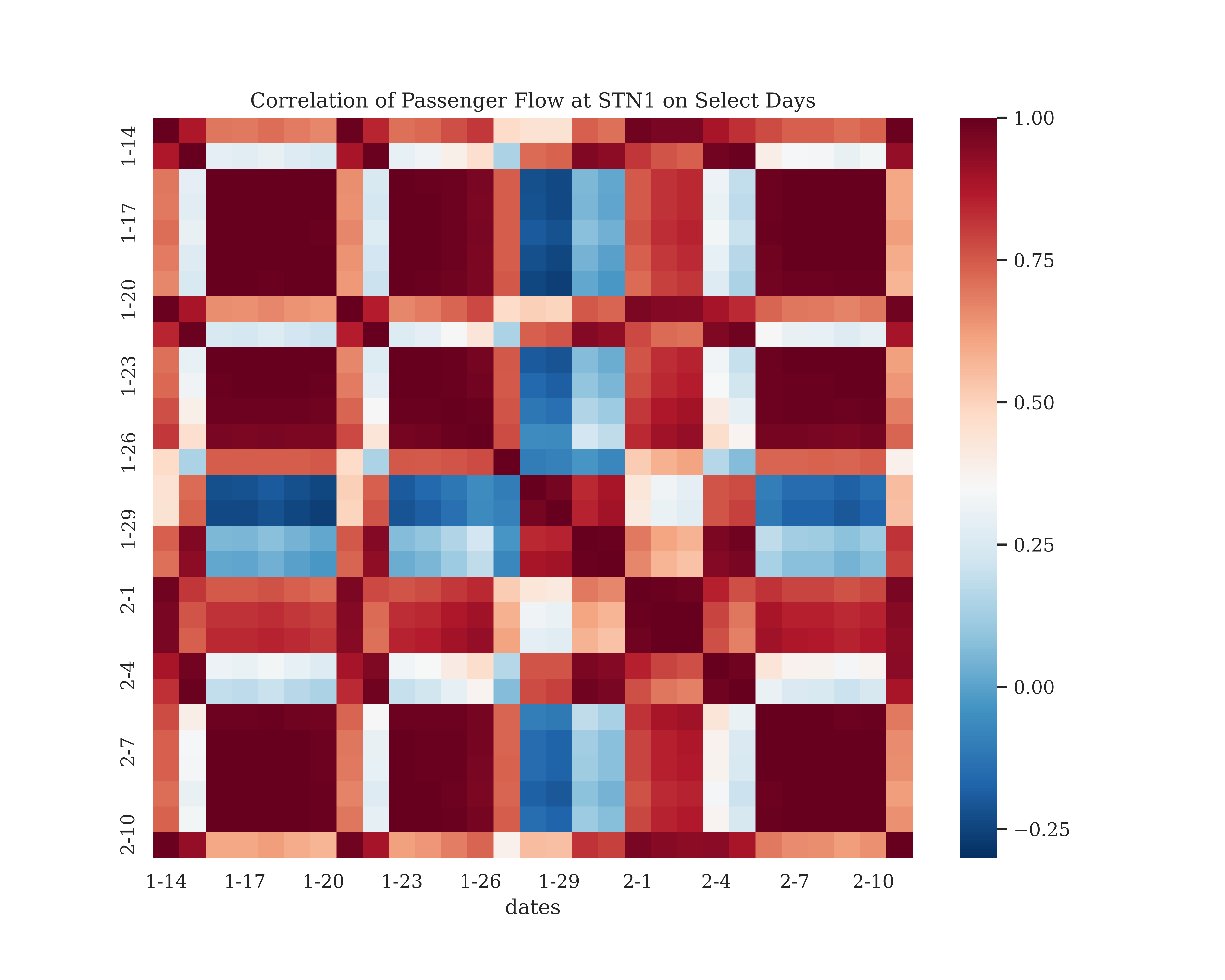}
    \includegraphics[width=0.9\linewidth]{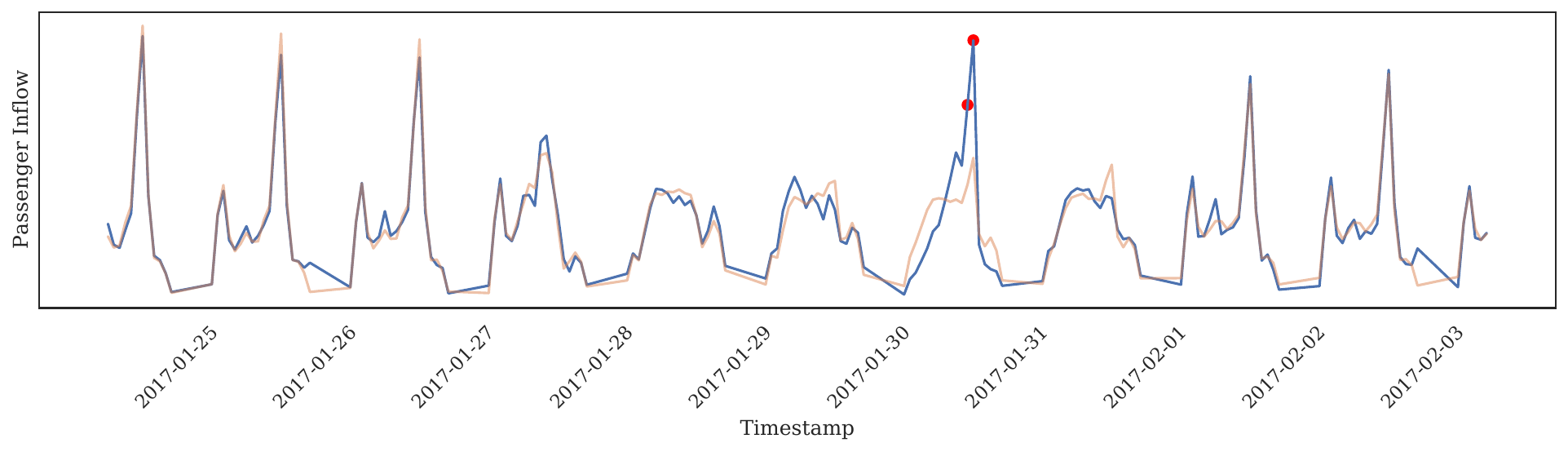}

    \caption{Correlation heatmaps showing (Left) geographical and contextual similarity between stations; (Right) temporal structure and sparse anomaly. (Bottom): Existence of sparse anomalies } 
    
    \label{fig:CorrelationHeatmaps}
\end{figure}

\begin{enumerate}
	\item \textbf{Complex spatio-temporal structure in high-dimensional data.} The dimensionality of the tensor data is often high (especially in the temporal domain), and traditional statistical methods may suffer from the “curse of dimensionality”. Furthermore, data points in the high-dimensional tensor often present complicated and high-order variations. For the temporal dependencies, passenger flow patterns show a strong periodicity with a period of 7 days, which denotes the weekly transit patterns \citep{li2020long}. This can also be seen in \Cref{fig:CorrelationHeatmaps} (right). Furthermore, on the same day, there are also complicated dependencies. For example, there are often two in-flow peaks for the multi-functional regions in the daily profile during the weekdays, which represent the morning and afternoon time for the daily commute, whereas a dominant morning in-flow peak is usually observed in residential regions and an evening in-flow peak commonly in business regions. {\color{black}For spatial structures, we create a correlation heatmap of passenger flows in different stations on day 1, as shown in \Cref{fig:CorrelationHeatmaps} (left). \Cref{fig:CorrelationHeatmaps} (left) demonstrates two kinds of  passenger flow similarities. The geographical similarity is based on the distance between the stations; while the contextual similarity is based on the function of the stations.}
	\item \textbf{Existence of sparse anomalies.} The tensor may contain sparse outliers due to measurement errors or rare events in urban passenger flow, such as weather, special occasions, etc. The sparse outliers may refer to the sudden increase of demand; this may be related to some unprecedented events happening, such as concerts and festivals, or this could even be due to maintenance or breakdowns in other stations whose spillover effect flows into the station in question. Identifying these events can also help prepare for the sudden increase in demand. {\color{black}To better understand the sparse anomaly feature, we create a correlation heatmap of passenger flows at station 1 on different days, as shown in \Cref{fig:CorrelationHeatmaps} (right). \Cref{fig:CorrelationHeatmaps} (right) shows that in most days, the passenger flow follows the temporal structure of the similarities of the weekday and weekend, except for 4 days. These days serve as the sparse anomaly in the data. Finally, to further illustrate the existence of "sparse anomalies", we have included another plot on the sparse anomalies, as shown in \Cref{fig:CorrelationHeatmaps} (bottom). A good example is illustrated here of an unexpected peak in the afternoon of January 30th, 2017, at the Fo Tan station. The peak is due to people leaving home from the Lunar New Year horse racing event, which takes place on the race course accessible from the Fo Tan station.} {\color{black} Therefore, in this research, we define the ``anomaly" study in this research as the entry-level pieces of the tensor data that do not fit the underlying distribution of the main features of the data. }

\end{enumerate}


In the literature, most of the methods implement a three-step approach for tensor clustering. First, tensor decomposition can be used to reduce the tensor dimensionality. Then, sparse outliers can be removed by robust tensor decomposition, and lastly, clusters can be identified by existing clustering algorithms such as K-Means or subspace clustering. In the first step of tensor decomposition, related to dimension reduction techniques, non-negative tensor decomposition \citep{Lin2018}, Tucker decomposition \citep{kolda2009tensor}, and CP decomposition \citep{kolda2009tensor} are widely used. In the step of anomaly detection, to handle sparse outliers, \cite{Hu2021} proposed a robust tensor decomposition method that can handle outliers data while achieving dimension reduction. Unlike standard robust tensor decomposition applications, they used the $L_{2,1}$ norm as they assumed fiber-sparse anomalies. Finally, in the step of clustering, it is quite common to apply clustering algorithms to the reduced-dimensional space. For example, \cite{Yang2019120688} proposed a hybrid approach that combines tensor decomposition with spectral clustering in a low-dimensional space.


However, none of these aforementioned works considers the interrelated nature of clustering, outlier removal, and dimension reduction \citep{aggarwal2015outlier}, with the interrelation threefold as follows:
\begin{enumerate}
\item \textbf{Dimension Reduction and Clustering:} Clustering directly in the high dimensions will suffer from the ``curse of dimensionality'', given that data becomes much more sparse and Euclidean distance is not a good distance measure in high-dimensional space. Therefore, dimension reduction techniques are often needed to project data in a low-dimensional space before clustering methods are applied \citep{Yang2019120688}.
    \item \textbf{Outlier Decomposition and Clustering:} On the other hand, if outliers on stations are not correctly identified or decomposed, the results of clustering for the overall data from the normal patterns will also be compromised.
    \item \textbf{Dimension Reduction and Outlier Decomposition:} Furthermore, a direct dimension reduction on the outliers-corrupted data will also lead to the introduction of erroneous information into the lower-dimensional products, i.e., decomposed latent matrices, which will hurt the downstream task performance (i.e., clustering) using the low-dimensional space.
\end{enumerate}

In conclusion, we claim that the tensor clustering problem, dimension reduction, and outlier decomposition problem are inter-correlated tasks and tied to each other, and isolated treatment of them may fail to provide satisfactory performance.


In this paper, we propose a \textbf{L}ow-Rank \textbf{R}obust \textbf{T}ensor \textbf{S}ubspace \textbf{D}ecomposition (LRTSD) technique to achieve dimensionality reduction, spatial clustering, and sparse anomaly decomposition \textit{simultaneously}. The proposed method is inspired by the recent development in the following three fields: Tucker decomposition \citep{kolda2009tensor}, tensor subspace clustering \citep{Vidal2011-ti, zhang2015low}, and sparse anomaly decomposition \citep{yan2017anomaly, li2022profile} (In this paper, we interchangeably use the term outlier and anomaly to represent the data which behaves outstandingly). Tucker decomposition is used as a dimensionality reduction technique, which reduces the original high-dimensional tensor data into a lower-dimensional core tensor while fitting orthonormal bases \citep{tucker1966some}. Furthermore, the proposed method is inspired by the sparse anomaly decomposition methods \citep{yan2017anomaly, li2022profile}, which aims to decompose the sparse anomaly component from the background component, with the $L_{1}$ norm applied to encourage the sparsity of detected anomalies. Finally, the proposed method is also inspired by the subspace clustering method \citep{Vidal2011-ti}, which has achieved especially good accuracy in high-dimensional clustering problems \citep{parsons2004subspace}. 

In this paper, we innovatively apply subspace clustering to the core tensor in the Tucker Decomposition, which is inherently more resilient to corruption/anomaly. To our knowledge, this is the first instance of such integration, and our simulation studies substantiate its superior performance.
The amalgamation of these three components significantly increases the complexity of the model. To address this, we have developed an efficient optimization algorithm to overcome this challenge, ensuring the generalizability, computational feasibility, and enhanced accuracy of the tensor clustering model.

The remainder of this paper is organized as follows. 
Section \ref{chap4:sec:literature} gives a brief overview of the basic tensor notations and multilinear algebra operation and subspace clustering. Section \ref{chap4:sec:methodology} provides a detailed introduction to the proposed subspace Tucker decomposition methods as well as an efficient optimization algorithm. Section \ref{chap4:sec:sim-study-results} conducts a simulation study, and Section \ref{chap4:sec:case-study-results} apply the proposed method to a real dataset based on Hong Kong Metro data. Section \ref{chap4:sec:conclusion} concludes and presents some future works.

\section{Literature Review}\label{chap4:sec:literature}
In this section, we will briefly review the methodology related to tensor decomposition and subspace clustering methods for tensor data. 
\subsection{Tensor Decomposition}
Related to robust tensor decomposition methods, most existing works apply the dimension reduction approach and tensor decomposition method for feature extraction.

Historically, all data is vectorized to the sample dimension, and the ordinary principal component analysis (PCA) is applied to the vectorized data to extract and monitor features, which is typically termed vectorized PCA (VPCA)  \citep{nomikos1994monitoring}. However, such a method sacrifices the detection power, since the vectorization on high-dimensional data destroys the innate cross-dimensional relations. 

Tensor is introduced as an efficient data structure to preserve inter-dimensional correlations well, such as complex spatiotemporal correlations among traffic data, as mentioned before. {\color{black} Tensor decomposition is developed accordingly to extract the features from the tensor data. Various tensor decomposition methods are developed, namely: (1) CANDECOMP/PARAFAC (CP) Decomposition \citep{hitchcock1927expression}, which represents a tensor $\mathcal{X} \in \mathbb{R}^{I_1 \times I_2 \times \dots \times I_K}$  as the weighted summation of a set of rank-one tensors; 
(2) Tucker Decomposition \citep{tucker1966some}, which decomposes a tensor into a core tensor $\mathcal{C} \in \mathbb{R} ^{J_1 \times J_2 \times \dots \times J_K}$ multiplied by a mode matrix $\mbU^{(k)} \in \mathbb{R}^{I_k \times J_k}$ along each dimension. The proposed method is based on the Tucker decomposition, and detailed operations of the Tucker decomposition will be introduced in Section \ref{subsec: tensor}.}

\cite{Yan2015-sh} proposed to utilize several anomaly detection and monitoring schemes based on Tucker Decomposition \citep{tucker1966some}, Multi-linear PCA \citep{lu2008mpca}, CP Decomposition \citep{hitchcock1927expression}, and Uncorrelated multi-linear PCA \citep{lu2008uncorrelated} with the application for image-based anomaly detection. However, such approaches do not consider the properties of sparse outliers for anomaly detection. To address this problem, tensor-based sparse anomaly decomposition techniques have been proposed \citep{fanaee2016tensor} and applied into video-imaging data \citep{Yan2018-ux}, urban traffic flow prediction \citep{sofuoglu2021low}, crime rate monitoring \citep{zhao2021rapid}, pandemic monitoring \citep{zhao2020rapid}, etc. However, the major limitation of these existing tensor-based anomaly decomposition methods is that they assume that the tensor follows a joint smooth and low-rank representation, which is not feasible for high-dimensional tensor data with multiple latent clusters behind the data distribution \citep{li2020tensor}. However, a direct application of the decomposition method to the tensor data with outliers for clustering will render an oversimplified model.
\subsection{Tensor Subspace Clustering}

On the other hand, there are some recent literature focusing on clustering methods for tensor data. \citep{drakopoulos2019tensor} gave an extensive methodology review of the tensor clustering. Some methods such as topic-model-based clustering \citep{ziyue2021tensor, li2022individualized}, dynamic tensor clustering \citep{sun2019dynamic}, multi-view subspace clustering \citep{zhang2015low}, dynamic subspace clustering \citep{zhang2020dynamic}, joint K-means and high-order SVD \citep{huang2008simultaneous} are proposed. 
Among them, subspace clustering has been a popular method for high-dimensional clustering \citep{parsons2004subspace}, which learns data representation
in certain low-dimensional subspaces and clusters of the data points. The subspace clustering is commonly formulated based on the data's self-expression property \citep{gao2015multi}, representing the original input data $\mbX$ by itself: $\mbX = \mbX \mbZ + \mbE$, where $\mbZ \in \mathbb{R}^{n \times n}$ is the subspace representation matrix, and the nonzero elements in $\mbZ$ correspond to the data points from the same subspace. Different properties could be achieved by introducing regularization terms into the representation matrix. More details will be explained in Section \ref{subsec: subspace}. However, like the other tensor clustering methods mentioned above, they are typically operated on the high-dimensional tensor directly and may suffer from the curse of dimensionality. For example, similar as tensor regression \citep{lock2018tensor, gahrooei2021multiple}, directly applying self-expression on high-dimensional data  
leads to an even higher-dimensional self-expressiveness term. 

Recently, there have been some works focusing on combining tensor low-rank decomposition and subspace clustering \citep{fu2014tensor, fu2016tensor}. Specifically,  \cite{fu2016tensor} proposed a subspace clustering that incorporates sparse dictionary learning into tucker decomposition, where the tucker core tensor could be the input tensor itself \citep{fu2014tensor} or the inverse matricization product of the dictionary and sparse representation \citep{fu2016tensor}. However, these subspace clustering methods are directly targeted at high-dimensional tensors, which are not suitable for tensors with sparse anomalies. Recently, there have been some efforts to combine subspace clustering methods with anomaly detection to handle the challenges of anomalies in multi-clustered datasets.

Recently, there have been a few developed tensor clustering methods based on deep learning and neural networks. \citep{wang2023tensorized} proposed the Tensorized Hypergraph Neural Networks (THNN) to learn the hypergraph structure on higher dimensions. In two-dimensional cases, further methods, such as spectral clustering, can be performed to cluster the nodes. However, this method only considered the hypergraph structure and did not take subspace learning and anomaly into consideration.
\citep{zhao2022multi} proposed the Reinforced Tensor Graph Neural Network (RTGNN) to learn multi-view graph data. This method contains three modules, which makes it time-consuming. Tucker decomposition only takes part in the initialization, and the anomaly is neglected in this method. 
\citep{huang2023model} extensively explored various methods, specifically on hyperspectral image clustering. The improvement achieved by neural network-based methods compared to tensor decomposition and self-regression-based methods is data-dependent. However, compared to the tensor-based approach, neural networks tend to become time-consuming and require many training samples. 

\subsection{Sparse Anomaly Decomposition}

Anomaly is usually detected based on the assumption that anomalous data do not conform to expected behavior with outstandingly different features from the homogeneous background \citep{chandola2009anomaly}. Traditional methods detect the anomaly based on distance and density, which assume that anomalies lie far away from the background or locate in a less dense area. Various definitions of distance or density have been introduced \citep{du2014discriminative}. For example, \citep{sun2022tensor} enhanced the CP decomposition by a deep CP decomposition neural network for feature extraction with an anomaly detection step. Nonetheless, to further achieve the separation of the anomalies, a further step such as clustering is still necessary. In this case,
a one-step decomposition is favored in practice. 

This paper is also motivated by the recent development of the decomposition-based methods \citep{sofuoglu2021low, yan2017anomaly, li2022profile}, which try to decompose the original input data into several components directly. For example, \cite{yan2017anomaly} decomposed input data into the smooth background and sparse anomaly, with smoothness regularization on the background and sparsity penalty on the anomaly. Later, the extensions to spatiotemporal data \citep{Yan2018-ux}, deep learning-based decomposition models \citep{zhao2022deep} are proposed. For example, \citep{li2022profile} considered more components, namely, smooth background, sparse anomaly, sample-specific deviation, and random noise, and they introduced a transfer learning mechanism \citep{pan2009survey} to solve the cold-start problem for anomaly decomposition. However, these methods are not developed for tensor data, and the decomposed components are pre-defined in a fixed manner. Recently, \cite{shen2022smooth} proposed a tensor-based decomposition method to decompose the static tensor background and smooth foreground. However, the method is designed for static background and cannot be directly used to learn the tensor data from different subspaces. 
Our method instead is to combine the sparse decomposition method with the tensor decomposition approach, and it learns the hidden components in a data-driven and unsupervised manner via tensor subspace.

To summarize, this work is the first that achi†eves dimensionality reduction, spatial clustering, and anomaly decomposition simultaneously. 

\section{Methodology}\label{chap4:sec:methodology}
In this section, we will first review the basic tensor algebra in Section \ref{subsec: tensor}. Then, we will give a brief review of the recent development of subspace clustering in Section \ref{subsec: subspace}.
More specifically, we will discuss the proposed formulation combining tensor subspace clustering in Section \ref{chap4:sec:formulation}. We will then discuss how to solve the formulation efficiently in Section \ref{sec: Optimization}. 

\subsection{Basic Tensor Notation and Multilinear Algebra} \label{subsec: tensor}

In this section, we introduce basic notations, definitions, and operators
in multilinear (tensor) algebra that is used in this work. Throughout
the chapter, scalars are denoted by lowercase italic letters, e.g.,
$a$, vectors by lowercase boldface letters, e.g., $\mba$,
matrices by uppercase boldface letter, e.g., $\mbA$, and tensors
by calligraphic letters, e.g., $\mathcal{A}$. For example, an order-$K$
tensor is represented by $\mathcal{A}\in\mathbb{R}^{I_{1}\times\cdots\times I_{K}}$,
where $I_{k}$ represents the mode-$k$ dimension of $\mathcal{A}$.
The mode-$k$ product of a tensor $\mathcal{A}$ by a matrix $\mbV\in\mathbb{R}^{P_{k}\times I_{k}}$
is defined as follows:
 $$
 (\mathcal{A}\times_{k}\mbV)(i_{1},\cdots,i_{k-1},j_{k},i_{k+1},\cdots,i_{K})=\sum_{i_{k}}A(i_{1},\cdots,i_{k},\cdots,i_{K})V(j_{k},i_{k}).
 $$
The Frobenius norm of a tensor $\mathcal{A}$ can be defined as follows: 
$$
\|\mathcal{A}\|_{F}^{2}=\sum_{i_{1},\cdots,i_{K}}A(i_{1},\cdots,i_{k},\cdots,i_{K})^{2}.
$$
The $n$-mode unfold operator maps the tensor $\mathcal{A}$ into matrix
$\mbA_{(n)}$, where the columns of $\mbA_{(n)}$ are
the $n$-mode vectors of $\mathcal{A}$.

Tucker decomposition decomposes a tensor into a core tensor multiplied
by a matrix along each mode, formulated as follows:
\begin{equation*}
    \mathcal{A}=\mathcal{C}\times_{1}\mbU^{(1)}\times_{2}\mbU^{(2)}\cdots\times_{K}\mbU^{(K)},
\end{equation*}
where $\mathcal{C} \in \mathbb{R}^{J_1 \times J_2 \dots \times J_K}$ is the core tensor,  $\mbU^{(k)}$ is an orthogonal $I_{k}\times J_{k}$ matrix
and is a principal component mode-$k$. The definition of Kronecker
product is as follows: Suppose $\mbA\in\mathbb{R}^{m\times n}$
and $\mbB\in\mathbb{R}^{p\times q}$ are matrices, the Kronecker
product of these matrices, denoted by $\mbA\otimes\mbB$,
is an $mp\times nq$ block matrix, which can be formalized as follows:
 $$
 \mbA\otimes\mbB=\left[\begin{array}{ccc}
	a_{11}\mbB & \cdots & a_{1n}\mbB\\
	\vdots & \ddots & \vdots\\
	a_{m1}\mbB & \cdots & a_{mn}\mbB
\end{array}\right]
$$









\subsection{Subspace Clustering} \label{subsec: subspace}

\begin{figure}
	\begin{centering}
		\includegraphics[width=0.7\textwidth]{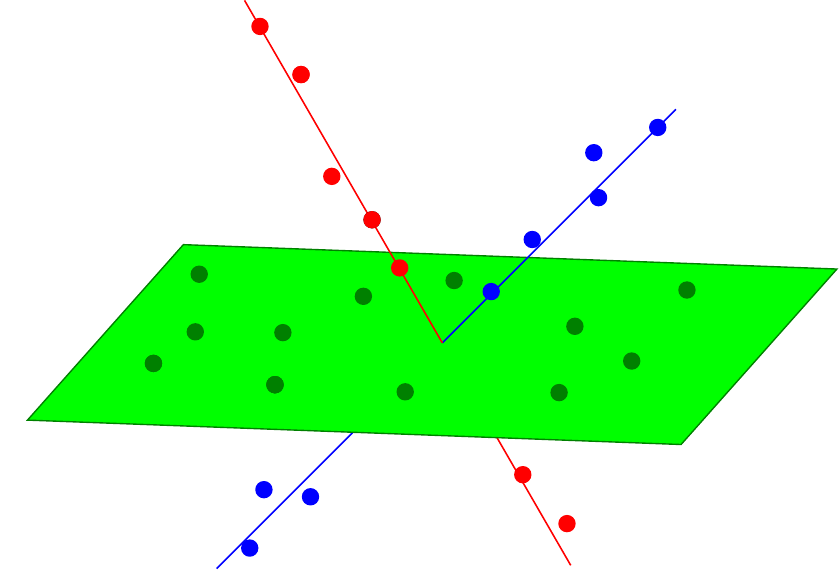}
		\par\end{centering}
	\caption{Data were drawn from two one-dimensional subspaces (red and blue)
		and one two-dimensional subspace (green). Adapted from \citep{Vidal2011-ti}.\label{fig:subspace-clustering-toy-example}}
\end{figure}

An important but often unnoticed assumption of PCA is that the data
lives in a \emph{single} low-dimensional subspace. This assumption
may be too crude for many applications, including spatiotemporal problems
such as motion segmentation \citep{vidal2008multiframe}. It is
more likely in such cases that the data lives in a mixture of subspaces
instead of a single one (see \Cref{fig:subspace-clustering-toy-example}).
The field of subspace clustering deals with problems of this nature
and has attracted growing interest in recent years \citep{liu2019robust,guo2019low}.

The central problem of subspace clustering is the grouping of points
and their respective clusters are both unknown in the beginning. Various
frameworks have been proposed in recent years, ranging from direct
algebraic solutions to probabilistic approaches \citep{Vidal2011-ti},
whereas spectral clustering \citep{Von_Luxburg2007-sb} dominates
the approaches proposed in recent years due to their superior performance
over other frameworks. In short, spectral clustering aims to build
a similarity graph where nodes are individual samples and the edges
are similarity links. Algorithms such as graph cut can be used later
to identify the clusters.

An effective way to obtain similarity is to use the self-expressiveness
property \citep{parsons2004subspace}. Let $\mbD\in\R^{n\times p}$ be a data matrix. The self-expressiveness can be defined with the equality $\mbD=\mbZ\mbD$ where
the self-expression matrix $\mbZ\in\R^{n\times n}$ helps to express
each point as a linear combination of other points in the data. This
matrix can later be used to find clusters, as explained above.

Spectral subspace clustering methods differ in how they
regulate the affinity matrix to avoid the trivial solution of identity
matrix $\mbZ=\mbI$. Sparse subspace clustering (SSC) imposes sparsity
on the affinity matrix \citep{yang2015sparse}, which is formulated
as follows:

\[
\text{\ensuremath{\underset{\mbZ}{\text{\text{min}}}}} \quad \gg\mbZ\gg_{1}+\frac{\lambda}{2}\gg\mbD-\mbZ\mbD\gg_{F}^{2}\quad s.t.\quad\text{\text{diag}}(\mbZ)=0,
\]

{\color{black}where the $\|.\|_1$ represent the L1 penalty of the self-expression matrix. By applying this penalty, the self-expression matrix $\mbZ$ is also assumed to be low-rank.
Low-rank representation method was proposed by \cite{Liu2013-tt}. Similarly,
 another method of low-rank subspace clustering was developed by
\cite{Vidal2014-fa} where the core formulation uses the nuclear-norm of the self-expression matrix $\mbZ$ defined by the sum of its singular values
as the best convex approximation of the rank of $\mbZ$.}
\[
\underset{\mbZ,\mbD,\mbA}{\text{min}} \quad \gg\mbZ\gg_{*}+\frac{\tau}{2}\gg\mbA-\mbA\mbZ\gg_F^2+\frac{\alpha}{2}\gg \mbD-\mbA\gg_F^2\quad s.t.\quad\mbZ^{\top}=\mbZ
\]

Thresholding ridge regression (TRR) \citep{peng2015robust} is an
improvement on SSC in the sense that it relaxes the
requirement to assume the structure of the affinity matrix, whether
it is low-rank or sparse. The formulation is also much simpler.
\[
\underset{\mbZ}{\text{min}} \quad \lambda\gg\mbZ\gg^{2}_{F}+\gg\mbD-\mbD\mbZ\gg_{F}^{2}
\]
The regularization parameter $\lambda$ represents the trade-off between
the reconstruction fidelity and clustering ability, which may be set
with respect to the use case. However, all these variations focus
on the matrix formulation and cannot be applied to the tensor formulation.

\subsection{Proposed Low-Rank Robust Tensor Subspace Decomposition (LRTSD) Model}
\label{chap4:sec:formulation}

Without loss of generality, we will illustrate the proposed method using a three-dimensional tensor.
The methodology can be easily extended into higher-order tensors. For a three-dimensional tensor, we assume that $\mathcal{X}\in\R^{N\times I_{2}\times I_{3}}$ to represent the original data, where the first dimension denotes the sample dimension that requires the clustering. Furthermore, we assume that this tensor $\mathcal{X}$ can be decomposed into the normal variations data $\mathcal{L}$ and the outlier tensor $\mathcal{A}$ (i.e., represents anomalous inflows in the passenger flow data), which is assumed to be sparse. We assumed the normal variations data $\mathcal{L}$ has a low-rank structure, which is a common assumption for high-dimensional tensors for dimension reductions. Furthermore, without loss of generality, we assume the first mode of tensor $\mathcal X$ 
is the dimension that needs further clustering. We name the first mode of the tensor (mode 1) as the clustering mode and the other modes (modes 2 and 3) as the non-clustering modes. 

For example, the passenger flow data can be represented as a tensor $\mathcal{X}\in\R^{N\times I_{2}\times I_{3}}$, with the three modes representing the station (with $N$ stations), the day of the week (with $I_2=7$ days), and the 5-minute time intervals (with $I_3 = 247$ intervals from 5 am to 1:30 am) within each day for the period where the subway station is open for, respectively. Each element of the tensor $X_{i,d,t}$ represents the inflow into stations $i$ at day $d$ and time interval $t$. Here, the station $i$ is the clustering mode, and $d$ and $t$ are the non-clustering modes, given that the station $s$ is often heterogeneous and may exhibit multiple clusters. 

In this case, we propose to use a partial Tucker decomposition on the non-clustering modes as:
\begin{equation}
    \mathcal{L}=\mathcal{C}\times_{2}\mbU_{2}\times_{3}\mbU_{3},
\end{equation}
where a core tensor $\mathcal{C}\in\R^{N\times P_{2}\times P_{3}}$ reduces the dimensionality of the original tensor. $\mbU_{2}$ and $\mbU_{3}$ are respective orthogonal matrices for the non-clustering
modes. We propose to find subspaces in the latent space on the core tensor $\mathcal{C}$
as opposed to original space, as this is computationally more efficient. Here, the notations that will be used in the paper are given in \Cref{tab:notation}.

\begin{table}[h]
	\caption{Notations for Proposed LRTSD Model.}
	\centering
	\begin{tabular}{l|p{70mm}}\toprule
		Notation & Explanation \\\midrule
		$\mathcal{X} \in \mathbb{R}^{N\times I_2 \times I_3}$ & $N$: number of samples, $I_2, I_3:$ feature dimensions \\
		$\mathcal{C} \in \mathbb{R}^{N\times P_2\times P_3}$ & Core tensor, with reduced dimensions $P_2, P_3$ \\
        $\mbU_l \in \mathbb{R}^{I_l,P_l}, l=2,3$ & orthogonal matrices for dimension reduction \\
        $\mathcal{A} \in \mathbb{R}^{N\times I_2\times I_3}$ & Anomaly tensor, same size as $\mathcal{X}$\\
        $\mbZ \in \mathbb{R}^{N\times N}$ & Self-expression matrix\\
		\bottomrule
	\end{tabular}
	\label{tab:notation}
\end{table}

Using the thresholding ridge regression framework \citep{peng2015robust}, we utilize the self-expressive expression $\mathcal{C}-\mathcal{C}\times_{1}\mbZ$ by
regularizing the self-expression matrix $\mbZ \in \mathbb{R}^{N \times N}$ using Frobenius norm. Formally, the model is described as an optimization problem:

\begin{align}
	\underset{\mbZ,\mathcal{C},\mbU_{2},\mbU_{3},\mathcal{A}}{\text{min}} &  & \begin{aligned}\frac{1}{2}\gg\mbZ\gg_{F}^{2}+\frac{\lambda_{z}}{2}\gg \mathcal{C}-\mathcal{C}\times_{1}\mbZ\gg_{F}^{2}+\lambda_{a}\gg\mathcal{A}\gg_{1}\\
		+\frac{\lambda_e}{2}\gg\mathcal{X}-\mathcal{A}-\mathcal{C}\times_{2}\mbU_{2}\times_{3}\mbU_{3}\gg_{F}^{2}
	\label{eq:mtr1-main-optim}
	\end{aligned}
\\
	&  & \mbU_{2}^{\top}\mbU_{2}=\mbI\label{eq:mtr1-main-u2tu2}\\
	&  & \mbU_{3}^{\top}\mbU_{3}=\mbI\label{eq:mtr1-main-u3tu3}
\end{align}

In summary, each term in the Equation (\ref{eq:mtr1-main-optim}) achieves following purpose:
\begin{itemize}
    \item $\gg\mathcal{X}-\mathcal{A}-\mathcal{C}\times_{2}\mbU_{2}\times_{3}\mbU_{3}\gg_{F}^{2}$ is the error term between the original input data $\mathcal{X}$ and the two decomposed components, i.e., anomaly $\mathcal{A}$ and low rank structure $\mathcal{L} = \mathcal{C}\times_{2}\mbU_{2}\times_{3}\mbU_{3}$;
    \item $\gg \mathcal{C}-\mathcal{C}\times_{1}\mbZ\gg_{F}^{2}$ is the self-expression term in tensor subspace clustering and is imposed on the decomposed core tensor. $\gg\mbZ\gg_{F}^{2}$ is the ridge regularization;
    \item $\gg\mathcal{A}\gg_{1}$ is the sparsity regularization term for the anomaly.
\end{itemize}
To this end, this model achieves what we promised ``a simultaneous framework for dimension reduction, spatial clustering, and anomaly detection''.

\subsection{Optimization Procedure\label{sec: Optimization}}
In this section, we propose an efficient algorithm to solve the proposed optimization problem \Cref{eq:mtr1-main-optim} - \Cref{eq:mtr1-main-u3tu3}.
We will utilize the Block Coordinate Descent procedure to obtain each one of the model parameters blocks $\mbZ,\mathcal{C},\mbU_{2},\mbU_{3}$ and $\mathcal{A}$ iteratively while keeping the others fixed from the last iteration. When \Cref{mtr:algo:LRSTDRRSC} terminates through convergence, graph cut is applied to $\mbL=\mbZ^{\top}+\mbZ$ to achieve clusters, and $\mathcal{A}$ can be analyzed for detected point anomalies. The rationale of the proofs of the propositions are relegated to the Appendix.

\begin{algorithm} 
  \caption{Low-rank and Sparse Tensor Decomposition with Ridge Regularized Subspace Clustering}
  \SetAlgoLined
  \KwData{$\mathcal{X}$,  $N$, $P_2$, $P_3$}
  \KwResult{$\mathcal{A}$, $\mathcal{C}$, $\mbU_2$, $\mbU_3$, $\mbZ$}
  Initialize $\mathcal{A}=\mathbf{0}$, $\mbZ=\mbI$. Initialize $\mathcal{C}$, $\mbU_2$, $\mbU_3$ by applying Tucker decomposition on $\mathcal{X}$\;
  \While{not converged}{
    Update $\mbZ$ as in \Cref{mtr:eqn:optimal-Z} \;
    Update $\mathcal{A}$ as in \Cref{mtr:eqn:optimal-A} \;
    Update $\mathcal{C}$ as in \Cref{mtr:eqn:optimal-C} \;
    Update $\mbU_2$ and $\mbU_3$ as in \Cref{mtr:eqn:optimal-U} \;
  }
  \label{mtr:algo:LRSTDRRSC}
\end{algorithm}

\subsubsection{Calculating the Self-expression Matrix}
The subproblem for optimizing $\mbZ$ when all other parameters are fixed is
\begin{align}
	\underset{\mbZ}{\text{min}} & \frac{1}{2}\gg\mbZ\gg_{F}^{2}+\frac{\lambda_{z}}{2}\gg\mbC_{(1)}-\mbZ\mbC_{(1)}\gg_{F}^{2}
	\label{mtr:eqn:subprob-Z}
\end{align}
\begin{proposition}\label{prop:update-Z}
The optimization problem in \Cref{mtr:eqn:subprob-Z} admits a closed-form solution,
\begin{align}
	\mbZ^{*} = \lambda_{z}(\mbI+\lambda_{z}\mbC_{(1)}\mbC_{(1)}^{\top})^{-1}(\mbC_{(1)}\mbC_{(1)}^{\top})\label{mtr:eqn:optimal-Z}
\end{align}
\end{proposition}






Equation \eqref{mtr:eqn:subprob-Z} suggests that there are only two terms in the problem related to the self expression matrix $\mbZ$. Proposition \ref{prop:update-Z} derived a closed-form solution.

\subsubsection{Calculating the Sparse Anomaly Tensor}
The subproblem for optimizing $\mathcal{A}$ when all other parameters are fixed is
\begin{align}
\underset{\mathcal{A}}{\text{min}}\lambda_{a}\gg\mathcal{A}\gg_{1}+\frac{\lambda_{e}}{2}\gg\mathcal{X}-\mathcal{A}-\mathcal{C}\times_{2}\mbU_{2}\times_{3}\mbU_{3}\gg_{F}^{2}
\label{mtr:eqn:subprob-A}
\end{align}

\begin{proposition}
The optimization problem in \Cref{mtr:eqn:subprob-A} admits a closed-form solution using the soft-thresholding operator,
\begin{equation}
	\mathcal{A}^{*}=\mathrm{sgn}(\mathcal{X}-\mathcal{C}\times_{2}\mbU_{2}\times_{3}\mbU_{3})\odot\max(\mbzero,|\mathcal{X}-\mathcal{C}\times_{2}\mbU_{2}\times_{3}\mbU_{3}|-\frac{\lambda_{a}}{\lambda_{e}})\label{mtr:eqn:optimal-A}
\end{equation}
\end{proposition}

The proof follows directly on using soft thresholding in solving the orthogonal LASSO problem and is therefore omitted here.

\subsubsection{Calculating the Core Tensor}
The subproblem for optimizing $\mathcal{C}$ is

\begin{align}
\underset{\mathcal{C}}{\text{min}}\frac{\lambda_{z}}{2}\gg \mathcal{C}-\mathcal{C}\times_{1}\mbZ\gg_{F}^{2}+\frac{\lambda_{e}}{2}\gg\mathcal{X}-\mathcal{A}-\mathcal{C}\times_{2}\mbU_{2}\times_{3}\mbU_{3}\gg_{F}^{2}
\label{mtr:eqn:subprob-C}
\end{align}

\begin{proposition}
Define $\mbM\triangleq\mbX_{(1)}-\mbA_{(1)}$ and $\mbP\triangleq(\mbU_{3}\otimes\mbU_{2})^{\top}$. The optimization problem in \Cref{mtr:eqn:subprob-C} admits a closed-form solution
\begin{align}
\mbC_{(1)}^{*} = \lambda_{e}(\lambda_{z}(\mbI-\mbZ)^{\top}(\mbI-\mbZ)+\lambda_{e}\mbI)^{-1}\mbM\mbP^{\top}
\label{mtr:eqn:optimal-C}
\end{align}
The resulting matrix $\mbC_{(1)}^{*}$ is folded back to the original tensor $\mathcal{C}$ after the calculation.
\end{proposition}






\subsubsection{Calculating the Orthonormal Bases}
When all other parameters are fixed, the subproblem to update $\mbU_{2}$ is

\begin{equation}
\begin{aligned}
	\underset{\mbU_{2}}{\text{min}} & \frac{\lambda_{e}}{2}\gg\mathcal{X}-\mathcal{A}-\mathcal{C}\times_{2}\mbU_{2}\times_{3}\mbU_{3}\gg_{F}^{2}\\
	& \mbU_{2}^{\top}\mbU_{2}=\mbI
	\label{mtr:eqn:subprob-U}
\end{aligned}    
\end{equation}

\begin{proposition}
Define $\mbQ\triangleq(\mbX_{(2)}-\mbA_{(2)})(\mbU_{3}\otimes\mbI)\mbC_{(2)}^{\top}$ and let $\mbQ=\hat{\mbU}\mbD\hat{\mbV}^{\top}$ be the singular value decomposition of $\mbQ$. Thus, the optimization problem in \Cref{mtr:eqn:subprob-U} admits a closed-form solution for the optimal values for the first orthogonal basis $\mbU_{2}^{*}$

\begin{align}
    \mbU_{2}^{*} = \hat{\mbU}\hat{\mbV}^{\top}
    \label{mtr:eqn:optimal-U}
\end{align}
\end{proposition}



Finally, the update of $\mbU_{3}^{*}$ is similar, therefore, omitted. In conclusion, the update of  $\mbU_{2}^{*}$ and $\mbU_{3}^{*}$ both yield closed-form solutions.





\section{Simulation Study \& Results}\label{chap4:sec:sim-study-results}

In this section, we will evaluate the proposed methods via a simulation study. We will first introduce the simulation setup in Section \ref{chap4:subsec:simulation-setup}. Then, the proposed method will be evaluated and compared with several clustering-based benchmark methods in Section \ref{SubSec:benchmark}.

\subsection{Simulation Setup}\label{chap4:subsec:simulation-setup}
In this section, we formalize the simulation procedure we use to generate clustered 3-order tensors with dimensionality $N\times I_2 \times I_3$. First, the procedure calls for the parameters in \Cref{chap4:tab:simulation-parameters}. After defining these parameters and setting the random number generating seed, we follow the following procedure to generate the data: 

\begin{itemize}
    \item {\color{black}For each cluster $k$ and mode of tensor $l=2,3$, generate random factor matrices $\mbU_{k,l} \in \mathbb{R}^{I_{l} \times P_{l}}$ from standard normal distribution, then orthogonalize $\mbU_{k,l}$ via the QR decomposition.}
    \item {\color{black}For each cluster $k$, generate a random core tensor $\mathcal{C}_k \in \mathbb{R}^{N \times P_{2}  \times P_{3}}$ from standard normal distributions.} \;
    \item For each cluster $k$, generate $\mathcal{X}_k  =  \mathcal{C}_k \times_{2} \mbU_{k1} \times_{3} \mbU_{k2} + \epsilon $, where noise is added on each element of $\mathcal{X}$ as $\epsilon \sim \mathcal{N}(0, \sigma^2)$, {\color{black} with $\sigma^2=0.25$}. \;
    \item Elementwise add sparse anomaly $\mathcal{X} \leftarrow \mathcal{X} + \psi \mathcal{A}_{p}$. Each entry of $\mathcal{A}_p$ is generated from the Bernoulli distribution with probability. Here $p$ is the anomaly ratio. {\color{black}In each entry where $\mathcal A_p$ is $1$, indicating anomaly is present, we add the anomaly intensity constant $\psi$ to the data.}
    \;
\end{itemize}

\begin{figure}
    \begin{centering}
       \includegraphics[width=0.7\linewidth]{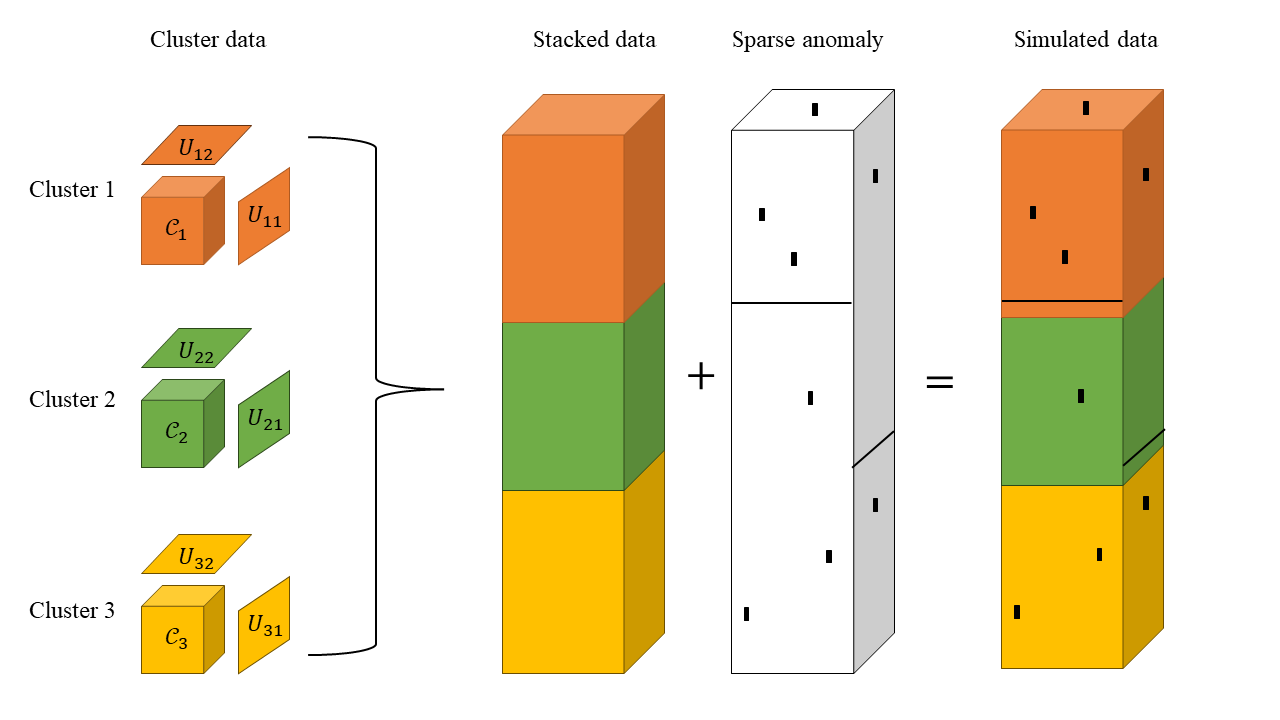} 
    \par\end{centering}
    \caption{Illustration depicting the generation process of simulated data}
    \label{fig:simulation_setup}
\end{figure}

  

\begin{table}[h]
	\caption{Required parameters for simulated clustered data generation and their notation.}
	\centering
	\begin{tabular}{l|p{70mm}}\toprule
		Notation & Explanation \\\midrule
		$K \in \mathbb{Z}^{+}$ & Number of clusters \\
		$N \in \mathbb{Z}^{+}$ & Number of samples per cluster \\
        $D_{l} \in \mathbb{Z}^{+}$ & Ambient dimension of $l$\textsuperscript{th} mode \\
        $P_{l} \in \mathbb{Z}^{+}$ & Intrinsic dimension of $l$\textsuperscript{th} mode \\
        $\sigma \in \mathbb{R}^{+}$ & Standard deviation of random noise \\
        $\mathcal{A}_p \in \mathbb{B}^{D_1\times \dots \times D_L}$ & Sparse anomaly index tensor with $0<p<1$ of entries being $1$ and others $0$\\
        $\psi \in \mathbb{R}$ & Sparse anomaly intensity \\
		\bottomrule
	\end{tabular}
	\label{chap4:tab:simulation-parameters}
\end{table}

{\color{black}\Cref{fig:simulation_setup} provides a visual representation of the simulation data generation process.} This procedure allows us to control key elements of the data generating process, such as the number of clusters, the cluster dimensions, the prevalence of anomalies, and how gross the anomalies are. We will now experiment with these factors and their effect on how well our model clusters samples and/or detects anomalies.

  

\subsection{Tuning Parameters Selection}\label{Subsec:Tuning}
{\color{black}
The proposed LRTSD has three tuning parameters: $\lambda_e$ penalizes the norm of the self-regression matrix; $\lambda_z$ penalizes the self-regression error of the core tensor; $\lambda_a$ penalizes the norm of the anomaly tensor. These three tuning parameters are in comparison to the reconstruction error term, which has a coefficient of 1. 

Related to the anomaly tuning parameter $\lambda_a$, one can select based on the percentage of nonzero anomaly entries expected in the dataset. For example, we can select the anomaly entries as a fixed percentage if we have some rough ideas on the percentage of the anomaly entries. If such a percentage is unknown, Otsu's method can be used to automatically select the tuning parameter $\lambda_a$, which aims to search for $\lambda_a$ that minimizes the intra-class variance, defined as a weighted sum of variances of the two classes: normal class and abnormal class. This procedure is discussed in detail in \citep{yan2017anomaly}. To select other tuning parameters, such as $\lambda_e$ and $\lambda_z$, we propose using the normalized cut score. For $\lambda_e$ and $\lambda_z$, $\lambda_e$ yields smaller diagonal values in the self-regression matrix $\bm Z$; higher $\lambda_z$ yields a more precise self-regression. To address the challenge of unknown clustering accuracy, we propose using the normalized cut to evaluate the affinity matrix produced by the algorithm \citep{zhang2023multi} and select the tuning parameters $\lambda_e$ and $\lambda_z$ that produce the most reasonable affinity matrix by minimizing the normalized cut score. This measurement is defined by $NC=\sum_{c=1}^C\frac{W_c^{out}}{W_c^{in}+W_c^{out}}$, where $c$ is the index of the cluster and $W_c^{in}$ and $W_c^{out}$ are the sum of the weights of the graph within the cluster and the weights of the graph outside the cluster. Smaller NC means better clustering.

\begin{figure}[t]
    \centering
    \includegraphics[width=0.48\linewidth]{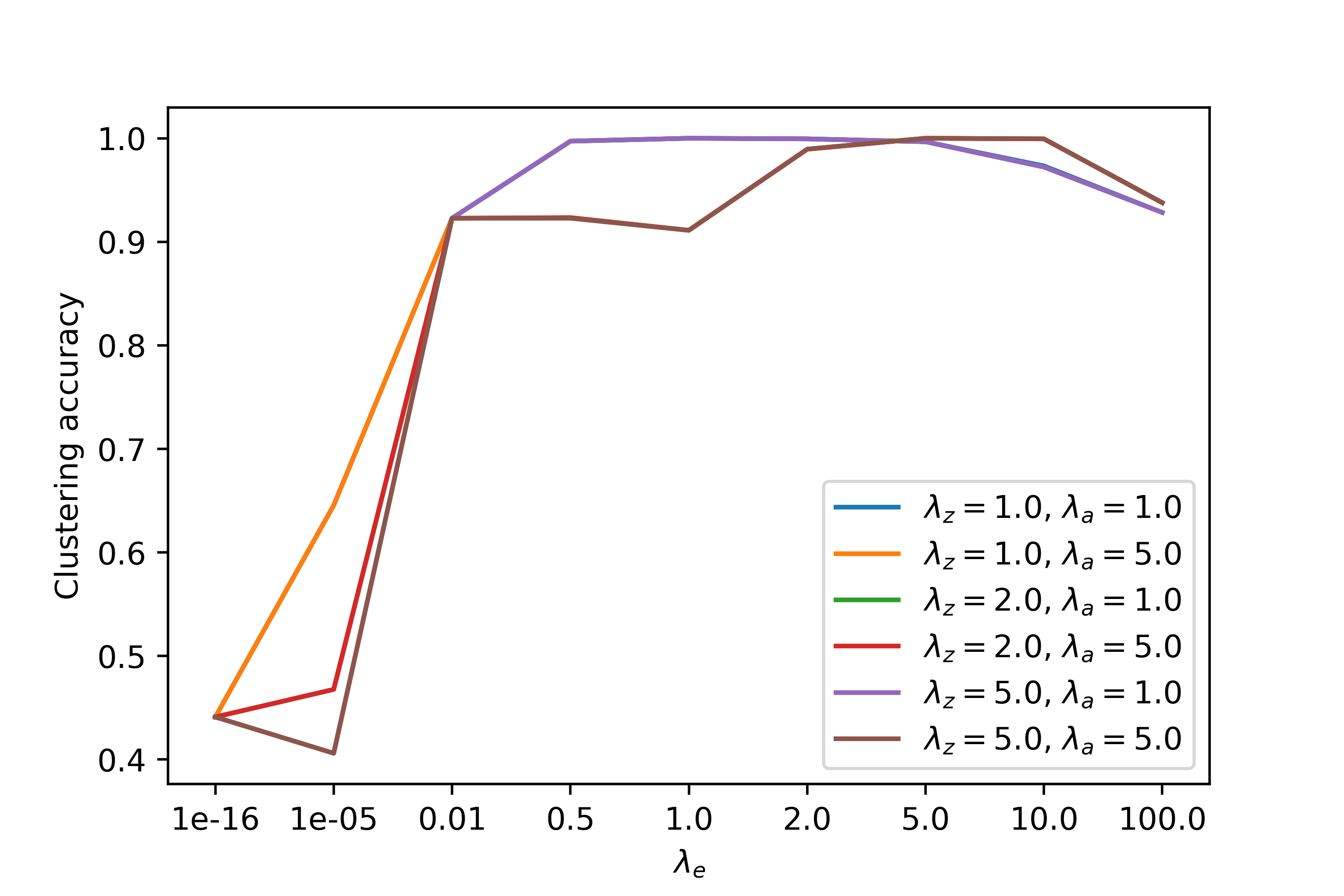}
    \includegraphics[width=0.48\linewidth]{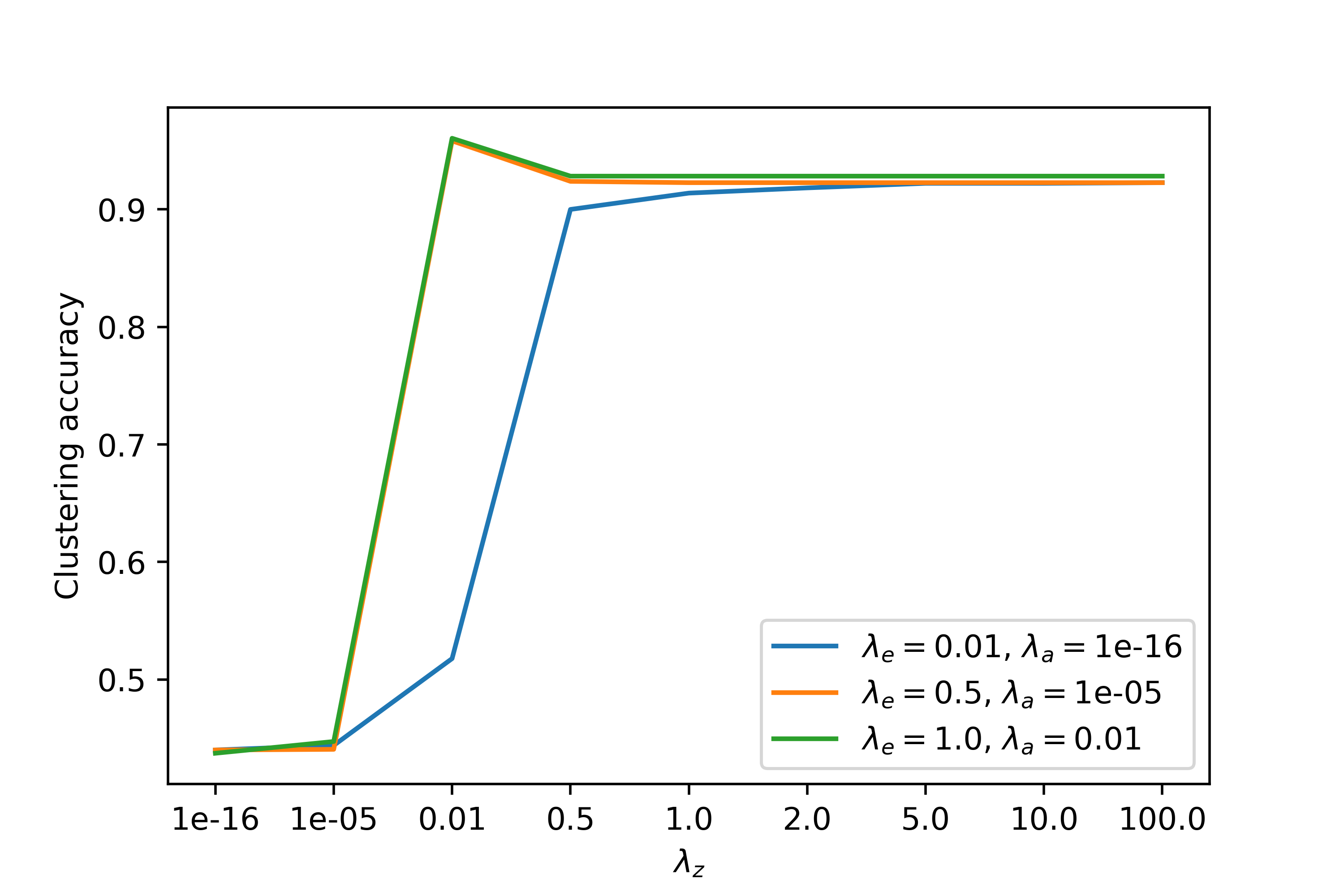}
    \includegraphics[width=0.48\linewidth]{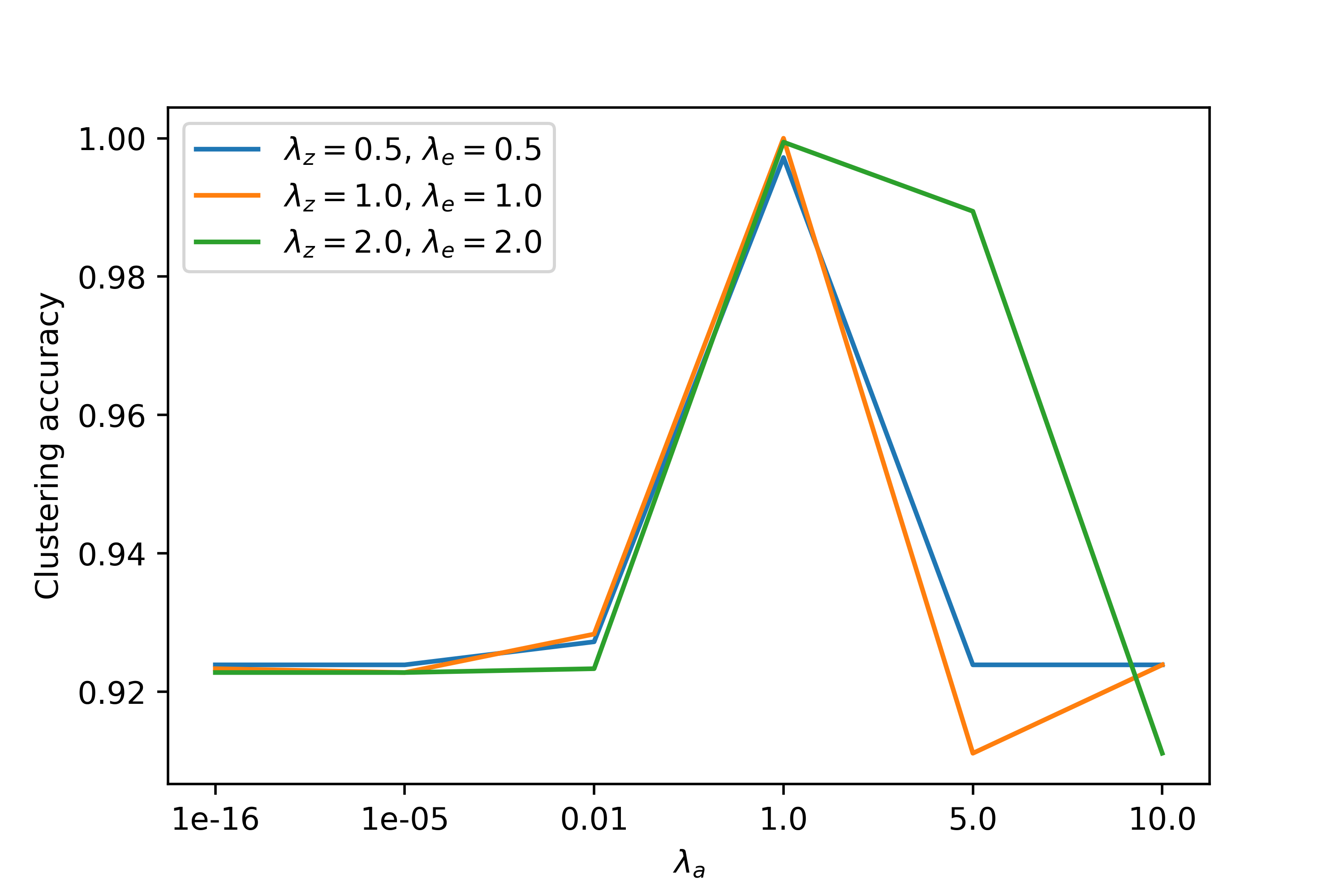}
    \caption{\color{black} Clustering Accuracy of Different Tuning Parameter Combinations}
    \label{fig:tuning}
\end{figure}

Some examples of clustering accuracy for different combinations of tuning parameters are shown in \Cref{fig:tuning}. From \Cref{fig:tuning}, we can see that (1) $\lambda_e$, $\lambda_z$, $\lambda_a$ cannot be too small, since we will almost lose a penalty term when any of the parameters are too small, which affects the clustering accuracy. (2) As $\lambda_e$ and $\lambda_z$ increase, the clustering precision increases. When $\lambda_e$ continues to increase, the clustering accuracy will decrease because the corresponding terms will dominate optimization. (3) $\lambda_a$ cannot be too small or too big, because we will not penalize the anomaly term when it is too small, and it will be too hard to discover any anomaly when $\lambda_a$ is too big.

}

\subsection{Performance Evaluation and Comparison}\label{SubSec:benchmark}
{\color{black}We generate a simulated three-dimensional tensor with three groups, thirty samples per cluster, fifty as the ambient dimensionality of each mode, and five as the intrinsic dimension on each mode.} Therefore, the dimensionality of the data tensor is $90\times50\times50$. This can be considered representative of a spatio-temporal observation with one spatial dimension and two temporal dimensions with the inherent clustered structure and redundant information with respect to dimensionality.
In this section, we compare the clustering accuracy of our models and three benchmark models under different anomaly ratios, and anomaly intensity, where anomaly ratios are defined as the ratio of anomaly entry compared to the entire tensor dimensions and anomaly intensity denotes $\psi$. The clustering accuracy is defined by the maximum proportion of matching labels of the $90$ objects in the first dimension over all permutations of the labels. The benchmark methods we compared are introduced below:
\begin{itemize}
    \item \textbf{K-means}: For each of the $90$ objects, we reshape the two-dimensional $50\times50$ matrix to a vector of length $2500$. We make $3$ groups using K-means based on the Euclidean distance.

    \item {\textbf{Robust Tucker+K means}: We first use the Robust Tucker decomposition introduced in \citep{heng2023robust, xue2017robust} to reduce the dimension to $15\times15$ for each object. We then use K means to generate $3$ clusters based on the vectorized core tensor.}

    \item \textbf{Multi-mode Tensor Space Clustering} \citep{he2022multi}: This is a method based on low-rank tensor representation (LTRR). We will use single-mode tensor space clustering for two reasons. Firstly, all the other methods use a single mode, and thus it makes for a fair comparison. Secondly, when we increased the number of modes, the running time increased dramatically. 
    
    \item \textbf{Robust Subspace Clustering} \citep{peng2015robust}: This method is based on the thresholding ridge regression (TRR) and is compatible with the matrix data. Therefore, we first reshaped the $90\times 50\times 50$ tensor into the $90\times 2500$ matrix and made $3$ clusters.
\end{itemize}

{\color{black} When comparing the performance of LRTSD and the benchmark methods, we use a grid search of the tuning parameters and report the best clustering precision in different settings, namely $\lambda_e$, $\lambda_c$, $\lambda_a$ in LRTSD, representation rank $r$ in LTRR, balance and thresholding parameter $\lambda$ and $k$ in TRR.} We report the mean accuracy of 20 replications of data generation and clustering accuracy for each anomaly ratio \& anomaly intensity combination and each method. The clustering accuracy result is shown in Table \ref{tab:benchmark_comparison} with their standard deviation over the 20 replications in parentheses.  In each combination of anomaly intensity and anomaly ratio, the best clustering accuracy is highlighted in bold. {From the clustering accuracy result, we can see that the K-means and Robust Tucker + K means methods yield the two lowest accuracy proportion. This is because they are not designed spedifically for clustering, and suffer from the curse of dimensionality when vectorized.} When the anomaly ratio and intensity are not large, LRTSD, TRR, and LTRR have almost perfect clustering accuracy. As the intensity and ratio of the anomalies increase, the LTRR and TRR start to show errors. The threshold at which the two methods start to have incorrect clustering labels is smaller than that of LRTSD.
Meanwhile, although LRTSD achieves the highest clustering accuracy, it does not require much more time compared to benchmark methods. {The average processing times for K-means, Robust Tucker + K means, LTRR, TRR, and LRTSD are 0.06 second, 0.68 second, 0.08 second, 0.51 second, 0.71 second, respectively.} 


\begin{table}[t]
    \centering
    \caption{{Clustering accuracy for LRTSD and other three benchmarks}}
    
    \begin{tabular}{|c|c|c|c|c|c|c|c|c|}
    
    \hline
        \multicolumn{2}{|c|}{Ratio} & \multirow{2}{*}{0.05} & \multirow{2}{*}{0.1} & \multirow{2}{*}{0.125} & \multirow{2}{*}{0.15} & \multirow{2}{*}{0.175} & \multirow{2}{*}{0.2} & \multirow{2}{*}{0.225}\\
        \cline{1-2}
        Intensity & Method & ~ & ~ & ~ & ~ & ~ & ~ & ~\\
        \hline
        \multirow{8}{*}{4} & \multirow{2}{*}{K mean} & 0.49 & 0.45 & 0.45 & 0.47 & 0.47 & 0.45 & 0.45 \\
        ~ & ~ & (0.02) & (0.03) & (0.03) & (0.02) & (0.02) & (0.02) & (0.03)\\
        \cline{2-9}
        ~ & \multirow{2}{*}{{Robust Tucker+K means}} & 0.5 & 0.5 & 0.49 & 0.45 & 0.48 & 0.47 & 0.45\\
        ~ & ~ & (0.03) & (0.03) & (0.03) & (0.02) & (0.02) & (0.02) &(0.02)\\
        \cline{2-9}
        ~ & LTRR & \textbf{1}(0) & \textbf{1}(0)  & \textbf{1}(0)  & \textbf{1}(0)  & \textbf{1}(0)  & \textbf{1}(0)  & \textbf{1}(0) \\
        \cline{2-9}
        ~ & \multirow{2}{*}{TRR} & \textbf{1}  & \textbf{1} & \textbf{1} & 0.99 & 0.98 & 0.97 & 0.97\\
        ~ & ~ & (0) & (0) & (0) & (0.01) & (0.01) & (0.01) & (0.01)\\
        \cline{2-9}
        ~ & LRTSD & \textbf{1}(0) & \textbf{1}(0)  & \textbf{1}(0)  & \textbf{1}(0)  & \textbf{1}(0)  & \textbf{1}(0)  & \textbf{1}(0)\\
        \hline
        \multirow{9}{*}{5} & \multirow{2}{*}{K mean} & 0.47 & 0.44 & 0.46 & 0.44 & 0.46 & 0.44 & 0.44 \\
        ~ & ~ & (0.02) & (0.02) & (0.03) & (0.03) & (0.03) & (0.03) & (0.03)\\
        \cline{2-9}
        ~ & \multirow{2}{*}{{Robust Tucker+K means}} & 0.47 & 0.47 & 0.48 & 0.45 & 0.43 & 0.43 & 0.43\\
        ~ & ~ & (0.03) & (0.03) & (0.02) & (0.03) & (0.03) & (0.03) & (0.02)\\
        \cline{2-9}
        ~ & \multirow{2}{*}{LTRR} & \textbf{1} & \textbf{1} & \textbf{1} & \textbf{1} & 0.99 & 0.98 & 0.97\\
        ~ & ~ & (0) & (0) & (0) & (0) & (0.01) & (0.01) & (0.02)\\
        \cline{2-9}
        ~ & \multirow{2}{*}{TRR} & 0.99 & 0.99 & 0.98 & 0.96 & 0.99 & 0.96 & 0.98\\
        ~ & ~ & (0.01) & (0.01) & (0.01) & (0.02) & (0.01) & (0.02) & (0.01)\\
        \cline{2-9}
         ~ & LRTSD & \textbf{1}(0) & \textbf{1}(0)  & \textbf{1}(0)  & \textbf{1}(0)  & \textbf{1}(0)  & \textbf{1}(0)  & \textbf{1}(0)\\
        \hline
        \multirow{9}{*}{6} & \multirow{2}{*}{K mean} & 0.44 & 0.44 & 0.43 & 0.44 & 0.43 & 0.42 & 0.42 \\
        ~ & ~ & (0.02) & (0.02) & (0.03) & (0.03) & (0.03) & (0.04) & (0.03)\\
        \cline{2-9}
        ~ & \multirow{2}{*}{{Robust Tucker+K means}} & 0.46 & 0.46 & 0.45 & 0.43 & 0.41 & 0.41 & 0.4\\
        ~ & ~ & (0.03) & (0.02) & (0.04) & (0.02) & (0.02) & (0.03) & (0.02)\\
        \cline{2-9}
        ~ & \multirow{2}{*}{LTRR} & \textbf{1} & \textbf{1} & 0.99 & 0.99 & 0.96 & 0.94 & 0.87\\
        ~ & ~ & (0) & (0) & (0.01) & (0.01) & (0.02) & (0.02) & (0.05)\\
        \cline{2-9}
        ~ & \multirow{2}{*}{TRR} & 0.99 & 0.96 & 0.97 & 0.96 & 0.9 & 0.87 & 0.81\\
        ~ & ~ & (0.01) & (0.02) & (0.02) & (0.02) & (0.04) & (0.05) & (0.05)\\
        \cline{2-9}
        ~ & {LRTSD} &\textbf{1}(0) & \textbf{1}(0)  & \textbf{1}(0)  & \textbf{1}(0)  & \textbf{1}(0)  & \textbf{1}(0)  & \textbf{1}(0)\\
        \hline
        \multirow{9}{*}{7} & \multirow{2}{*}{K mean} & 0.44 & 0.41 & 0.4 & 0.41 & 0.4 & 0.39 & 0.41 \\
        ~ & ~ & (0.02) & (0.03) & (0.03) & (0.03) & (0.04) & (0.03) & (0.03)\\
        \cline{2-9}
        ~ & \multirow{2}{*}{{Robust Tucker+K means}} & 0.45 & 0.43 & 0.43 & 0.4 & 0.39 & 0.4 & 0.39\\
        ~ & ~ & (0.04) & (0.03) & (0.03) & (0.04) & (0.02) & (0.03) & (0.02)\\
        \cline{2-9}
        ~ & \multirow{2}{*}{LTRR} & \textbf{1} & 0.99 & 0.98 & 0.96 & 0.83 & 0.74 & 0.64\\
        ~ & ~ & (0) & (0.01) & (0.01) & (0.02) & (0.07) & (0.09) & (0.12)\\
        \cline{2-9}
        ~ & \multirow{2}{*}{TRR} & 0.97 & 0.94 & 0.9 & 0.79 & 0.76 & 0.64 & 0.63\\
        ~ & ~ & (0.01) & (0.02) & (0.04) & (0.1) & (0.09) & (0.11) & (0.08)\\
        \cline{2-9}
        ~ & {LRTSD} & \textbf{1}(0) & \textbf{1}(0)  & \textbf{1}(0)  & \textbf{1}(0)  & \textbf{1}(0)  & \textbf{1}(0)  & \textbf{1}(0)\\
        \hline
        \multirow{10}{*}{8} & \multirow{2}{*}{K mean} & 0.44 & 0.39 & 0.39 & 0.4 & 0.39 & 0.41 & 0.4 \\
        ~ & ~ & (0.02) & (0.02) & (0.03) & (0.02) & (0.03) & (0.03) & (0.03)\\
        \cline{2-9}
        ~ & \multirow{2}{*}{{Robust Tucker+K means}} & 0.42 & 0.41 & 0.41 & 0.4 & 0.4 & 0.39 & 0.39\\
        ~ & ~ & (0.03) & (0.04) & (0.03) & (0.03) & (0.02) & (0.02) & (0.03)\\
        \cline{2-9}
        ~ & \multirow{2}{*}{LTRR} & \textbf{1} & 0.94 & 0.86 & 0.74 & 0.58 & 0.54 & 0.53\\
        ~ & ~ & (0) & (0.01) & (0.04) & (0.07) & (0.09) & (0.08) & (0.08)\\ 
        \cline{2-9}
        ~ & \multirow{2}{*}{TRR} & 0.98 & 0.86 & 0.82 & 0.78 & 0.74 & 0.51 & 0.47\\
        ~ & ~ & (0.01) & (0.05) & (0.08) & (0.11) & (0.09) & (0.06) & (0.07)\\
        \cline{2-9}
        ~ & \multirow{2}{*}{LRTSD} & \textbf{1}  & \textbf{1}  & \textbf{1}  & \textbf{1}  & \textbf{1}  & \textbf{1}  & \textbf{0.99}\\
        ~ & ~ & (0) & (0) & (0) & (0) & (0) & (0) & (0.01)\\
        \hline
    \end{tabular}
    \label{tab:benchmark_comparison}
\end{table}

\subsection{Sensitivity Analysis}
More specifically, we aim to answer two major questions by comparing the proposed methods with several benchmark methods in terms of anomaly detection accuracy and clustering accuracy. 
\begin{itemize}
	\item Question 1: \textit{If sparse and gross anomalies are present, does anomaly detection lead to better clustering?}
	\item Question 2: \textit{Does the tensor decomposition and dimension reduction lead to better clustering?}
\end{itemize}

\subsubsection{The Effect of Dimensionality Reduction on Clustering Accuracy}\label{chap4:subsec:the-effect-of-clusrtering-on-anomaly-detection}

We generate a simulated three-dimensional tensor with three clusters,  thirty samples per cluster, and 125 as the data dimensionality on each mode, namely, $90 \times 125 \times 125$, where the intrinsic dimensionality is $5$ in each cluster. Therefore, the total intrinsic dimensionality of the tensor is actually $5\times 3 = 15$. This can be considered representative of a spatiotemporal observation with one spatial dimension and two temporal dimensions with the inherent clustered structure and redundant information with respect to dimensionality. 

\begin{figure}
	\begin{centering}
		\includegraphics[width=0.8\textwidth, height=0.6\textwidth]{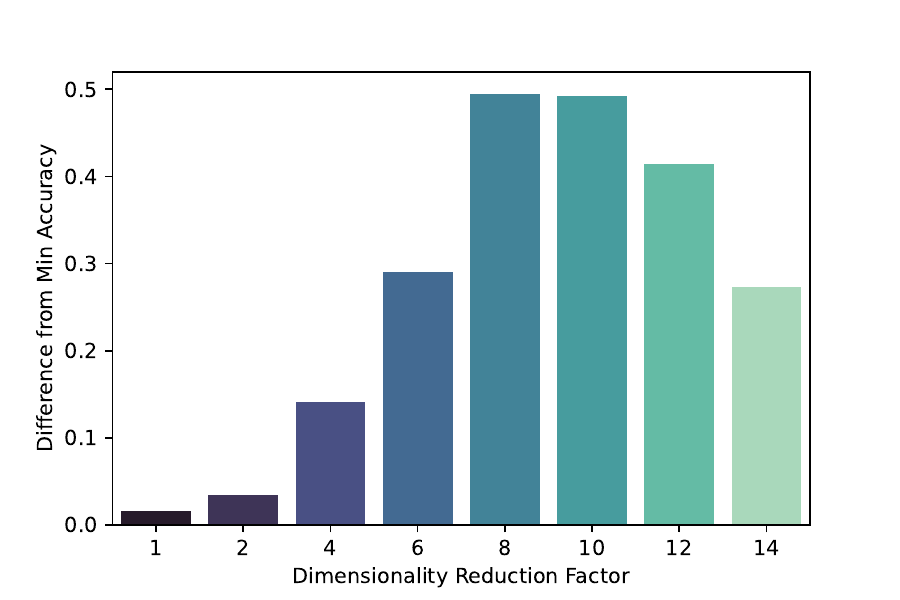}
		\par\end{centering}
	\caption{The evolution of clustering accuracy over multiple replications by increasing dimensionality factor on each mode. \label{fig:chap4-dimred-experiment}}
\end{figure}

For each seed, we generate some data and apply dimensionality reduction by a certain factor on each mode. The dimension reduction factor $\rho_d = [(\prod_{l} I_l/P_l)^{1/l}]$ is defined as the geometric mean of the data dimensionality $I_l$ over the intrinsic dimension $P_l$ averaged on different mode $l$ of the tensor.
For one factor of the experiment, we do not apply any dimensionality reduction (which is denoted by a factor of 1 on the $x$-axis in \Cref{fig:chap4-dimred-experiment}). In other words, we directly apply subspace clustering on the high-dimensional tensor, optimizing the following objective:
\[
\underset{\mbZ}{\text{min}} \quad \lambda\gg\mbZ\gg^{2}_{F}+\gg\mbX_{1}-\mbX_{1}\mbZ\gg_{F}^{2}
\]

\Cref{fig:chap4-dimred-experiment} shows the result of difference from minimum accuracy. For each replication, we obtain the clustering accuracy for different dimensionality reduction factors. Then they are subtracted from the minimum clustering accuracy within that replication and get the difference for each dimensionality reduction factor. After $20$ replications, we get the result of $20$ accuracy differences for each dimensionality reduction factor. Their mean is plotted in \Cref{fig:chap4-dimred-experiment}.
Observing \Cref{fig:chap4-dimred-experiment}, there is an optimal range of dimensionality reduction ratio around $8$ to $10$, which roughly corresponds to the true intrinsic dimension ratio, i.e., $125 / 15 = 8.33$. This clearly suggests that clustering is more effective at the right proportion of dimensionality reduction, supporting our claim that dimension reduction improves the tensor clustering performance. 

\subsubsection{The Effect of Anomaly Detection on Clustering Performance}\label{chap4:subsec:the-effect-of-anomaly-detection}
The parameters for generating the simulation data are set to be the same as those in Section \ref{SubSec:benchmark}. 

We factorize our experiment over the properties of the added sparse anomaly. Specifically, we alter how prevalent the anomaly is and how large its intensity is. At each factor of the experiment, we define the control group as the absence of anomaly detection, in other words, removing the term $\frac{\lambda_{a}}{2}\gg\mathcal{A}\gg_{1}$ from \Cref{eq:mtr1-main-optim}. The treatment group is \Cref{eq:mtr1-main-optim} as is; therefore, the algorithm tries to detect the sparse anomaly. We do this over multiple seeds, where in each seed the exact same observations are used for both the control and treatment groups. We report the difference in clustering accuracy between the treatment group and the control group to identify whether anomaly detection would increase clustering accuracy or not. The clustering accuracy is defined as almost the same as the well-known clustering accuracy, except that the permutation of the labels is allowed such that it maximizes the agreement between the predicted labels and the ground truth.

\begin{figure}
	\begin{centering}
		\includegraphics[width=0.8\textwidth]{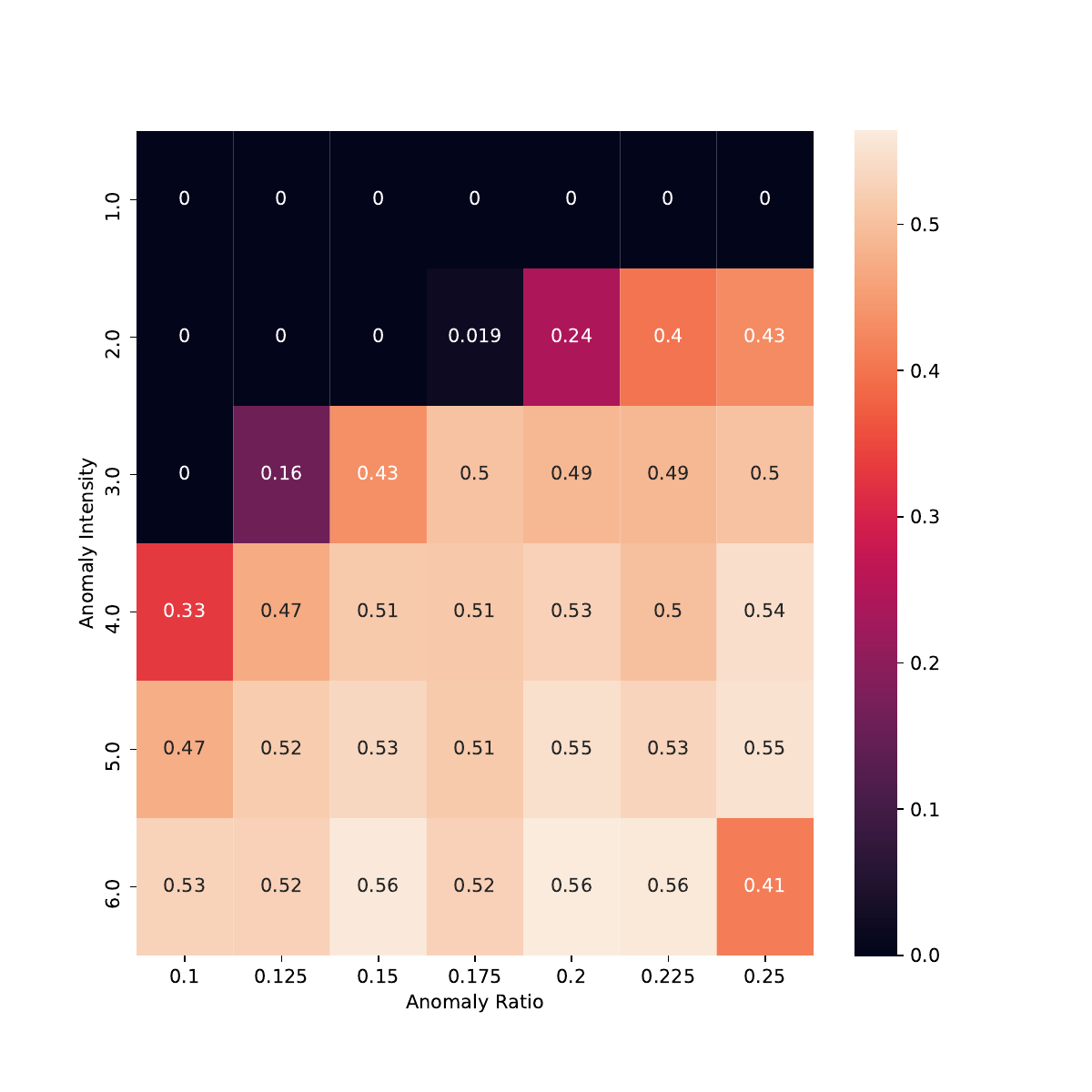}
		\par\end{centering}
	\caption{The average of pairwise differences in clustering accuracy between when anomaly detections is allowed and when it is not, across various anomaly conditions and ten data generating seeds. \label{fig:chap4-anomaly-detection-experiment}}
\end{figure}

In \Cref{fig:chap4-anomaly-detection-experiment}, we observe a clear relationship between the severity of the anomalies and how much anomaly detection improves clustering performance. As the intensity of the anomaly increases and/or the prevalence of anomalies increases, the application of anomaly detection becomes even more significant in terms of finding the right groupings between the samples of the data. This can be considered quite an expected finding, but it is an important sanity check of our optimization procedure, and it finalizes the discussion of why simultaneous anomaly detection is crucial for the quality of cluster discovery.

\section{Case Study \& Results}\label{chap4:sec:case-study-results}
We apply our proposed methodology to the smart card data collected by MTR, the company that operates Hong Kong's subway system. Smart card data emerged as an important ingredient for urban mobility analysis, as it can be collected cheaply and contains detailed information about how people move within a city \citep{pelletier2011smart}. Using this information, decisions at the tactical or strategic level can be made in a data-driven, thus more effective manner.

The data span an 82-day time horizon from January \nth{1} 2017 and record every tap-in and tap-out event from 85 stations of the system. We resample these events hourly for each day within the operating hours of the system. Thus, we end up with the number of people entering a station every operating hour of every station.

Such data can be represented as a three-dimensional tensor. The first mode of the tensor represents the stations, while the second and third modes represent the day and the time of day, respectively. Thus, we expect to obtain clusters at the station level using the clustering objective and point anomalies at hourly resolution. 

Understanding commonalities between stations is important from an urban mobility perspective. The stations that are similar to each other in terms of passenger inflow may reveal what kind of features about those stations are the most influential in people's decision to take the subway on a specific day and time of day.

Detecting anomalies is also important from a managerial point of view. Generally, urban mobility shows strong predictability due to the regular weekly schedules of city dwellers. However, rare events could occur, and the subway management would rather take precautions in the face of such events. Detection of past point anomalies may reveal the underlying reasons for one-off events that lead to extreme passenger inflow, such as concerts, celebrations, or extreme weather events. Armed with this knowledge, subway management can be more accurate in predicting future rare events.

\subsection{Station Clustering}
\begin{figure}
    \begin{centering}
        \includegraphics[width=0.5\linewidth]{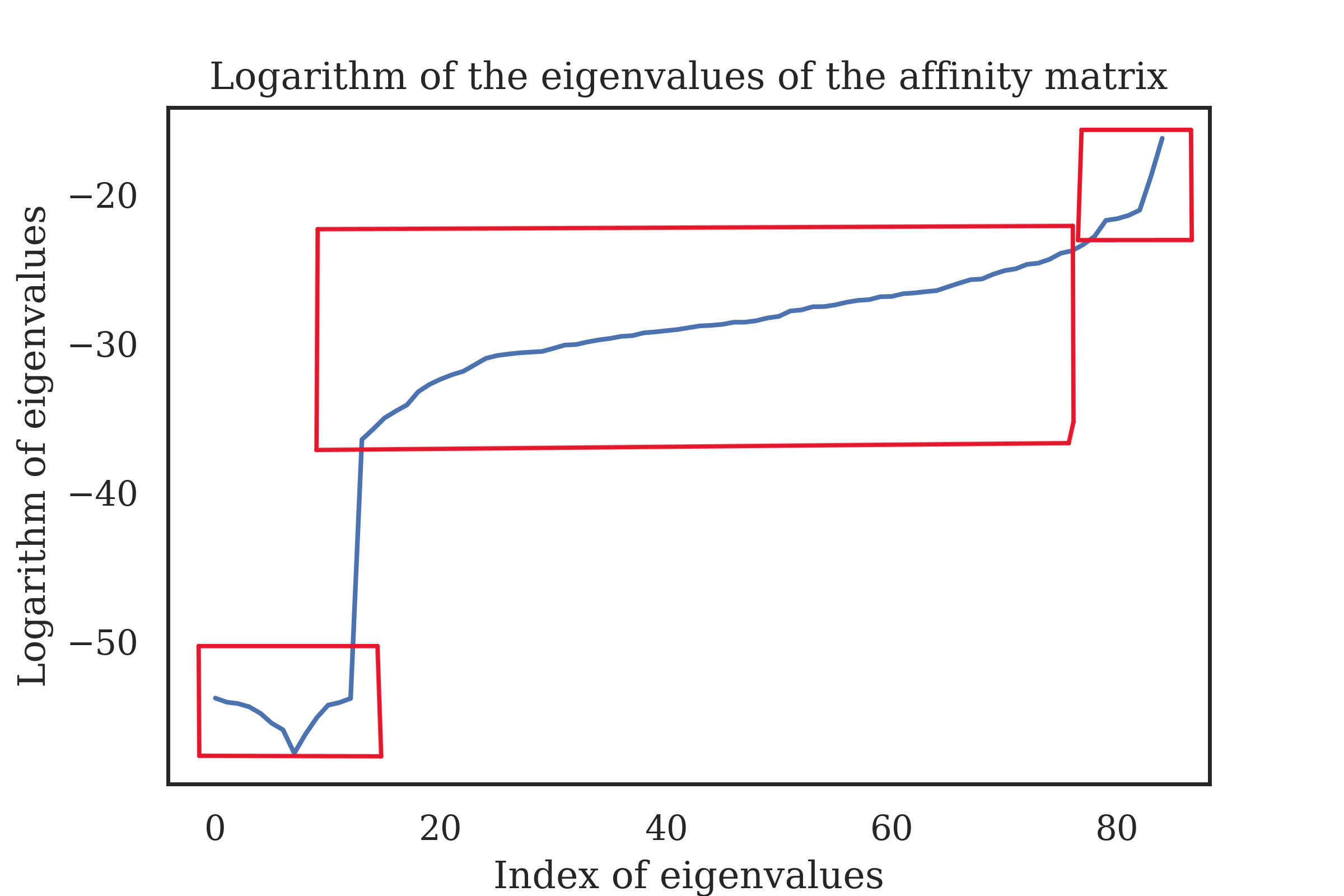}
    \par\end{centering}
    \caption{Eigenvalue spectrum of the affinity matrix to locate discontinuous segments}
    \label{fig:chap4-eigenvalues}
\end{figure}
We trained our model with the smart card data and recovered three clusters. The number of clusters was determined using the method introduced in \cite{von2007tutorial}. We first plot the logarithm of the eigenvalue spectrum of the Laplacian of the affinity matrix, then locate the discontinuous segments, as shown in Figure \ref{fig:chap4-eigenvalues}.

\begin{figure}
	\begin{centering}
		\includegraphics[width=0.9\textwidth]{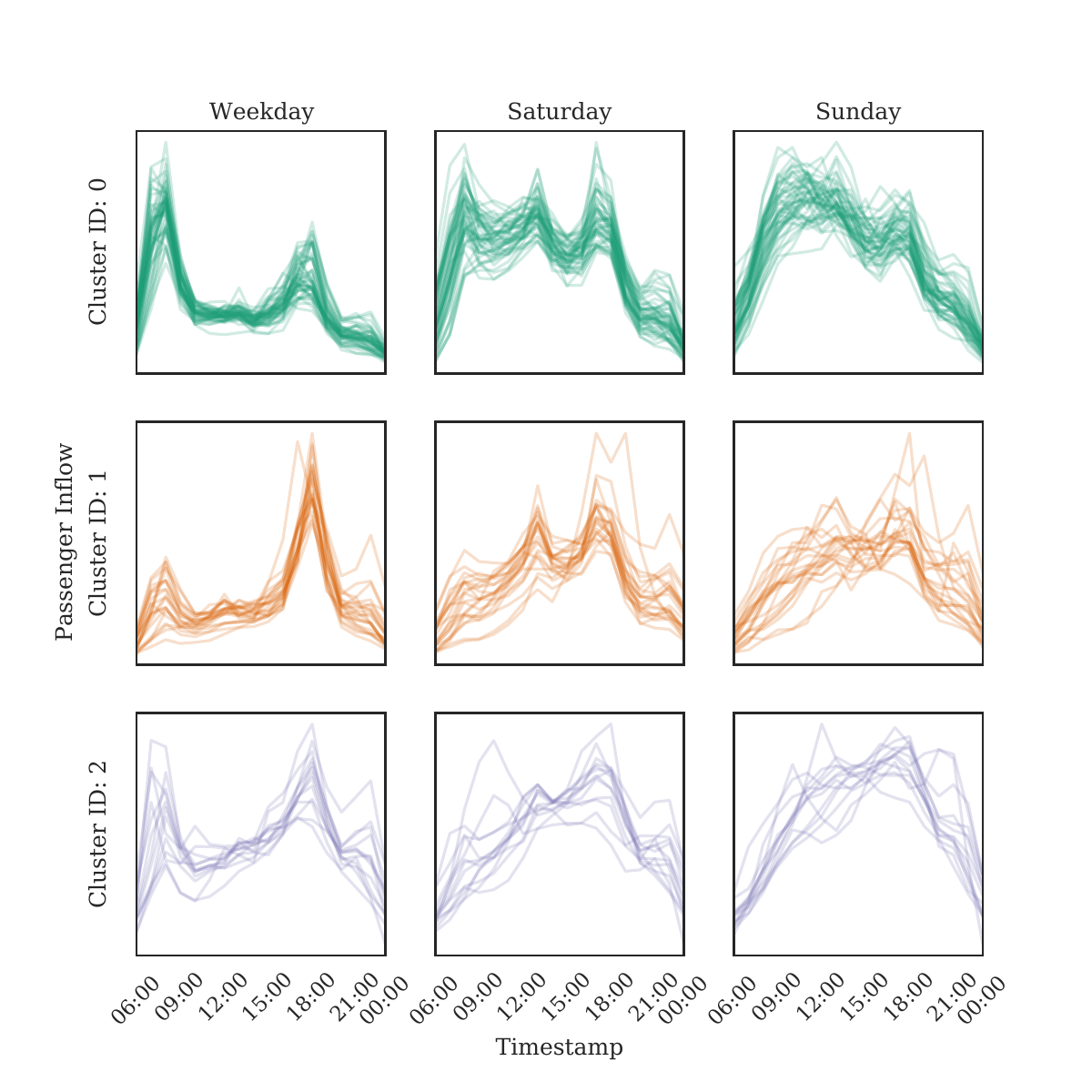}
		\par\end{centering}
	\caption{Hourly inflow patterns per cluster on a randomly picked week, across weekdays and weekend days. The color coding matches with the stations in \Cref{fig:chap4-three-way-clustering}. \label{fig:chap4-three-way-weekly-patterns}}
\end{figure}

\begin{figure}
	\begin{centering}
		\includegraphics[width=0.85\textwidth]{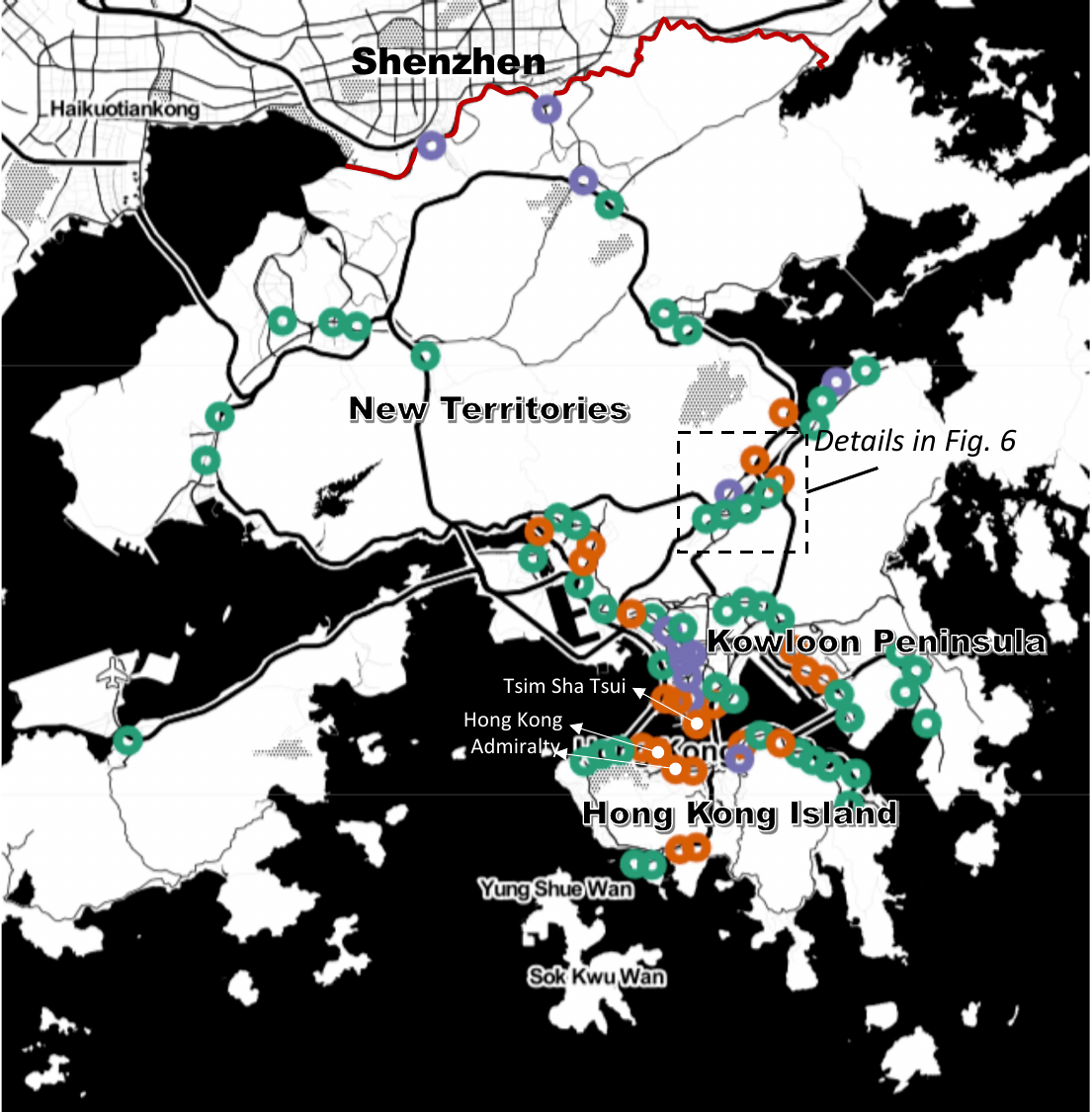}
		\par\end{centering}
	\caption{Hong Kong subway stations on a map, clustered three-way. Colors represent different clusters. Circle radius represents a typical walking distance of 400 meters around the station.  \label{fig:chap4-three-way-clustering}}
\end{figure}

\Cref{fig:chap4-three-way-weekly-patterns} and \Cref{fig:chap4-three-way-clustering}, when observed jointly, reveal important insight into how the stations are clustered. Weekdays play the primary role in determining the cluster of a station. As shown in the \Cref{fig:chap4-three-way-weekly-patterns}, the first cluster (Cluster ID: 0, colored in green) has a strong inflow peak at the morning peak hour.  These stations are also mainly located in the residential areas of Hong Kong (i.e., west and east end of Hong Kong Island, west side of New Territories, east side of Kowloon), as shown in the \Cref{fig:chap4-three-way-clustering}, Likewise, for the second cluster (Cluster ID: 1, colored in orange), we observe a strong evening peak inflow as opposed to the mornings, and they are mostly located in the business region such as the Central area of Hong Kong Island and the opposite side in Kowloon, as shown in \Cref{fig:chap4-three-way-clustering}. As we could conclude so far, for the most part, this can be explained by the land use, specifically, whether it is residential or work and office space. The third cluster (Cluster ID: 2, colored in purple) does not have a dominant peak between the morning and evening, and the rest of the inflow is rather evenly distributed throughout the day. While not as prominent as weekdays, weekends (especially Saturday) patterns also differ from one cluster to another.

Interpretation of the third cluster requires some contextual investigation. Hong Kong draws millions of visitors each year, especially from the bordering Shenzhen City, Guangdong Province of mainland China. The third cluster is a collection of stations where visitors enter Hong Kong, with the primary intention being cross-border shopping. The absence of taxes and duties on goods in Hong Kong makes it an appealing destination for Chinese visitors to shop. Since shopping is not a strictly time-bound activity, we observe inflows that are dispersed more homogeneously around the day, both on weekdays and weekends.

\begin{figure}
	\begin{centering}
		\includegraphics[width=0.85\textwidth]{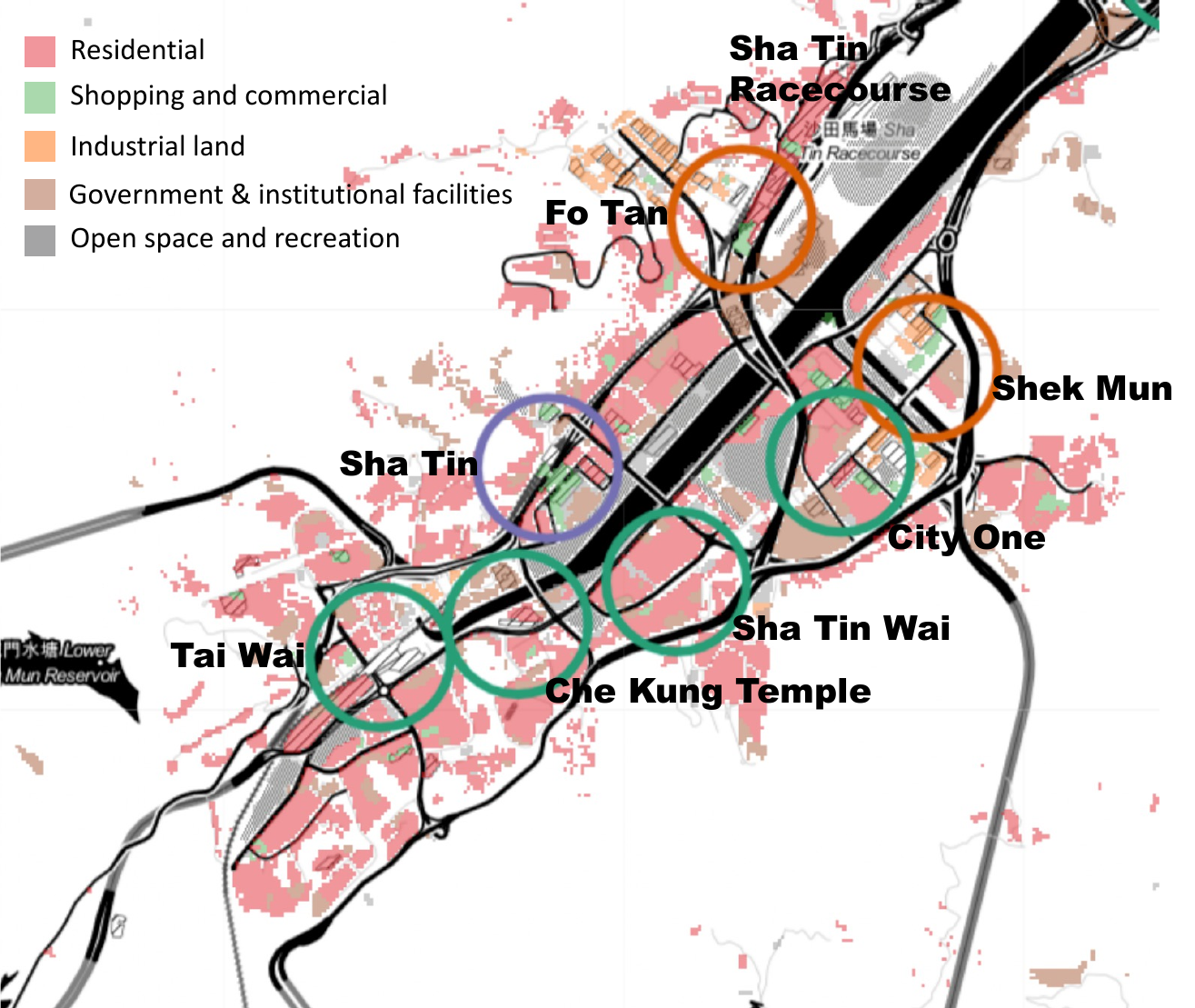}
		\par\end{centering}
	\caption{Sha Tin and Tai Wai regions with land use information. The circle color coding of clusters matches with the stations in \Cref{fig:chap4-three-way-clustering}. The land use color coding logic is as follows: \textbf{Red:} residential, \textbf{Green:} commercial, \textbf{Orange:} industrial, \textbf{Brown:} institutional, governmental, and \textbf{Grey:}  parks and recreations.  \label{fig:chap4-shatin-taiwai}}
\end{figure}

We now bring in the land use information \footnote{Open data via \url{https://www.pland.gov.hk/pland_en/info_serv/open_data/landu/index.html}} to the picture. In \Cref{fig:chap4-shatin-taiwai}, the first and second clusters can strictly be separated by the prevalence of residential areas or commercial and industrial areas, respectively. The only station from the third cluster, Sha Tin station, is surrounded by many shopping malls, with an iconic \textit{New Town Plaza} which offers around 2 million sq. ft. of exceptional more than 400 shopping, dining, and lifestyle facilities. Thus, it is a popular shopping destination for passengers visiting from mainland China. 

We can analyze even more thoroughly with granular information. The key takeaway is that the clustering of stations yields important insight into the factors that determine passenger inflows.

 	
 
 
 

\begin{figure}
 	\centering

    \begin{tabular}{@{}c@{}}
    \includegraphics[width=.8\linewidth]{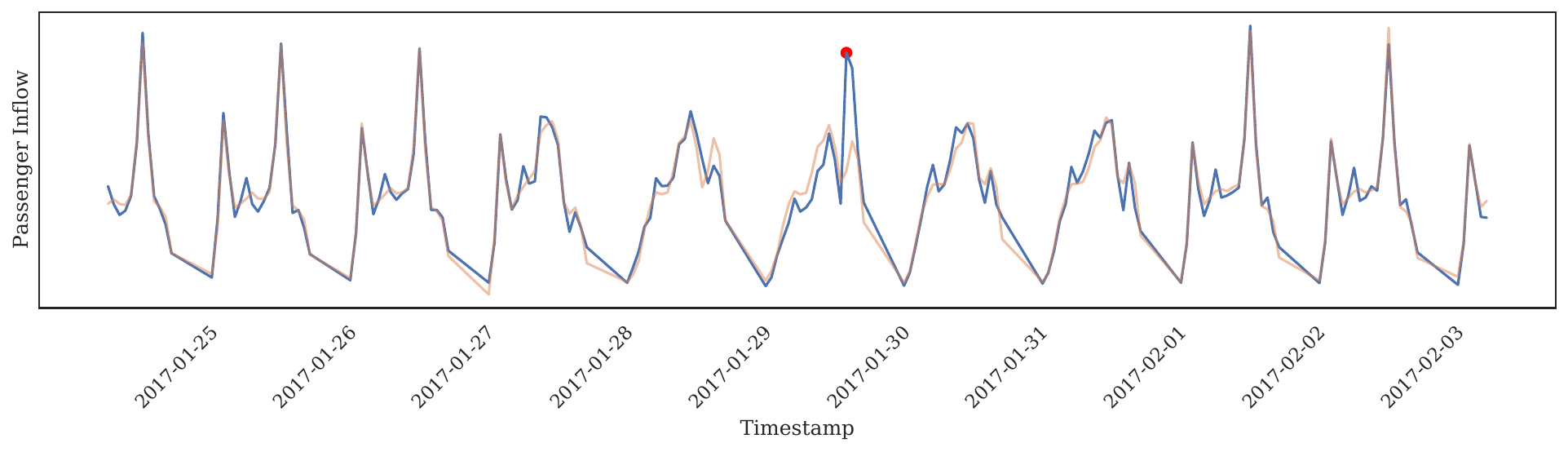} \\
    \small (a) Tsim Sha Tsui
  \end{tabular}

      \begin{tabular}{@{}c@{}}
    \includegraphics[width=.8\linewidth]{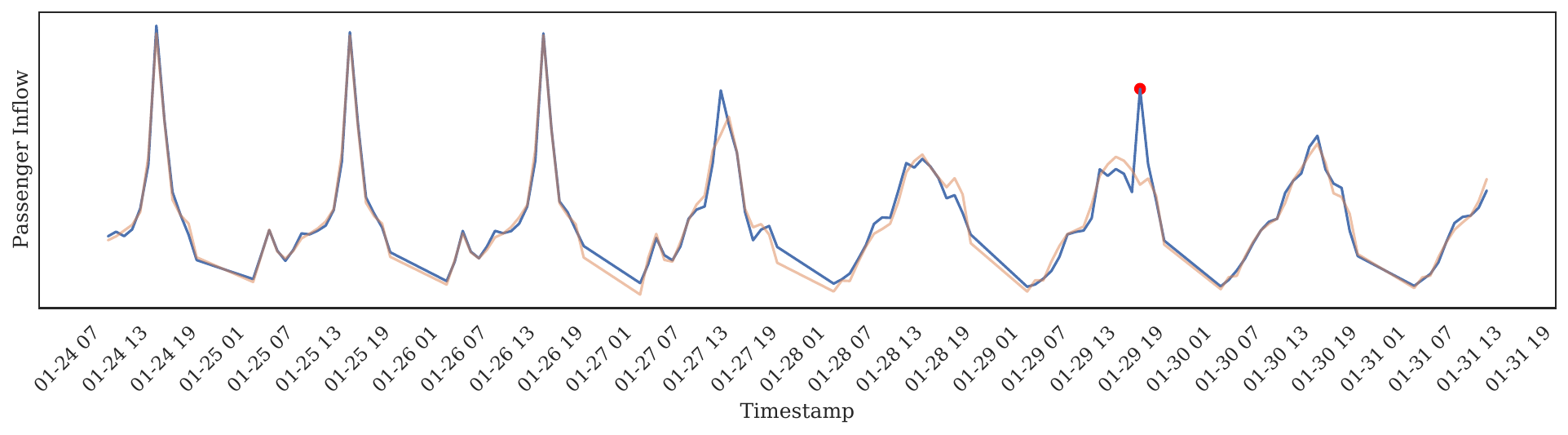} \\
    \small (b) Hong Kong
  \end{tabular}
  

       \begin{tabular}{@{}c@{}}
    \includegraphics[width=.8\linewidth]{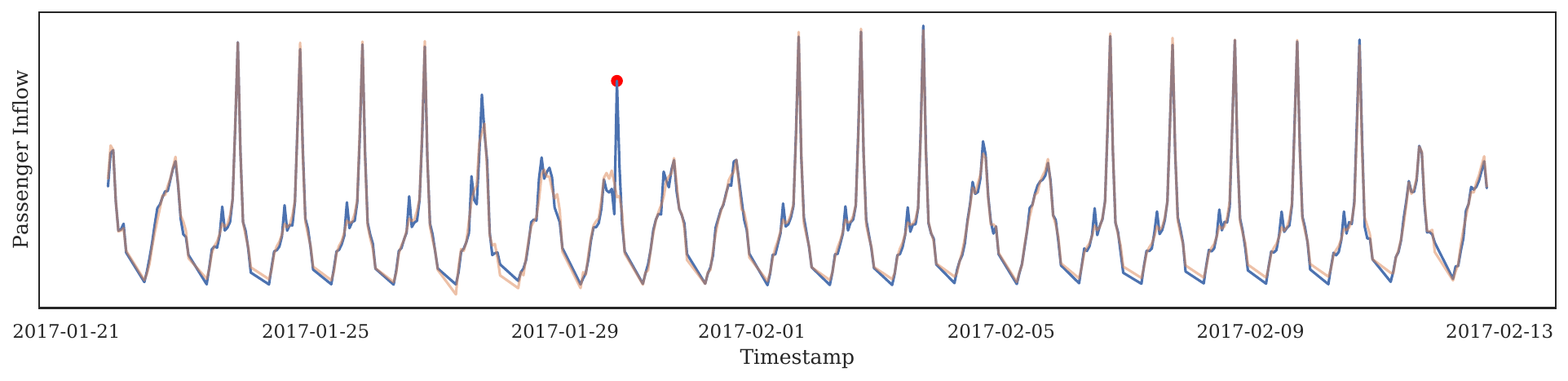} \\
    \small (c) Admiralty
  \end{tabular}
  

        \begin{tabular}{@{}c@{}}
    \includegraphics[width=.8\linewidth]{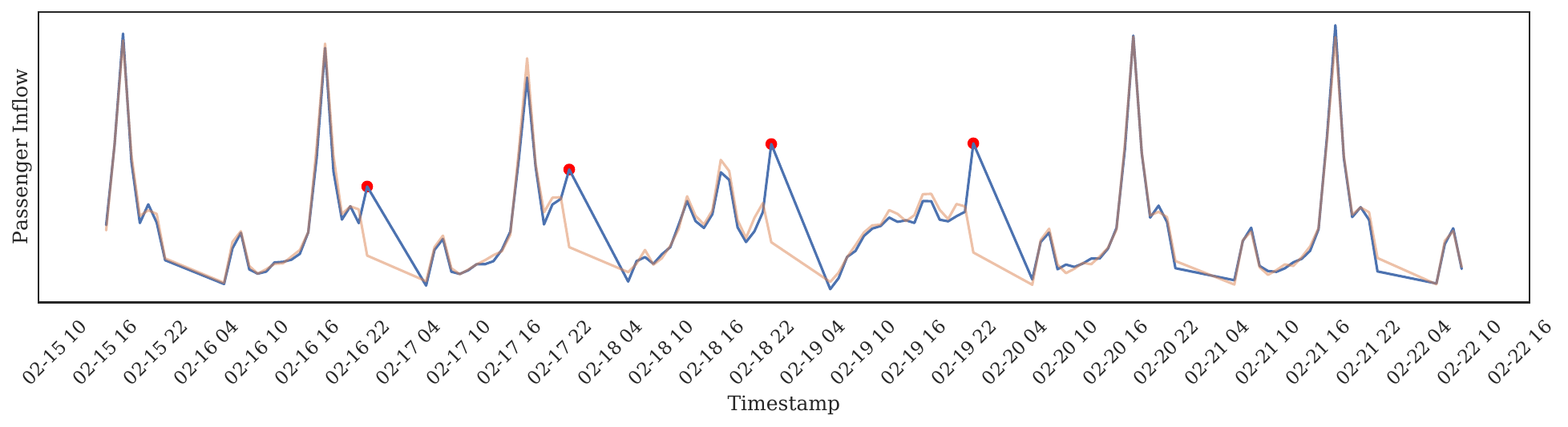} \\
    \small (d) Hung Hom
  \end{tabular}
  
 	\caption{Examples of rare events in five different stations, detected as point anomalies over passenger inflows.}
 	\label{fig:three graphs}
 \end{figure}

\subsection{Detected Point Anomalies}
We now focus on the point anomalies detected by the proposed model in the data from the smart card. As we have discussed before, one-off events are an important determinant of irregular patterns in passenger inflows. 


Lunar New Year celebrations in Hong Kong, which took place between \nth{28} and \nth{31} of January in Victoria Harbor between Hong Kong Island and the Kowloon Peninsula, yield a number of other rare events. One of the events is the Lunar New Year Fireworks Display. In \Cref{fig:three graphs}, we observe the peaks related to that event in the Tsim Sha Tsui station, Admiralty and Hong Kong, with locations pinned in \Cref{fig:chap4-three-way-clustering}. These stations are close to places that are considered the best places to watch the fireworks display.

Another interesting rare event takes place at the Hung Hom station. In \Cref{fig:three graphs}, we observed an unusual peak at 8 pm for four consecutive days from February \nth{16} to February \nth{19}.  This suggests the scheduled end of a four-day event, perhaps a conference, given the proximity of the station to Hong Kong Polytechnic University.

In general, detected anomalies are promising in terms of the amount of insight they can provide for public transportation management.

\section{Conclusion}\label{chap4:sec:conclusion}
In this paper, we focused on the problem of spatio-temporal modeling with tensors. Specifically, we proposed a tensor decomposition-based model that is capable of simultaneous clustering and point anomaly detection, all of which are important challenges to be addressed in this kind of data. Our simulation study shows that, in general, point anomaly detection increases clustering accuracy when point anomalies exist in data and vice versa. As for the case study, we trained our model on a collection of smart card transactions from the subway system of Hong Kong. Our model was able to effectively recover station clusters that are semantically consistent both temporally and spatially. The model was also able to recover rare events in terms of point anomalies, which offers valuable insights and assistance for managerial analysis and operations.

%
%
%
\begin{APPENDICES}

\section{Appendix}

    \subsection{Proof of Proposition 3.1}
        \begin{proof}
         We rearrange the objective functions as follows:

\begin{align*}
	\mathcal{L}_{\mbZ} & \triangleq\frac{1}{2}\gg\mbZ\gg_{F}^{2}+\frac{\lambda_{z}}{2}\gg\mbC_{(1)}-\mbZ\mbC_{(1)}\gg_{F}^{2}\\
	& =\frac{1}{2}tr(\mbZ^{\top}\mbZ)+\frac{\lambda_{z}}{2}tr(\mbC_{(1)}^{\top}\mbC_{(1)}-2\mbC_{(1)}^{\top}\mbZ\mbC_{(1)}+\mbC_{(1)}^{\top}\mbZ^{\top}\mbZ\mbC_{(1)})\\
	& =\frac{1}{2}tr(\mbZ^{\top}\mbZ)+\frac{\lambda_{z}}{2}(tr(\mbC_{(1)}^{\top}\mbC_{(1)})-2tr(\mbC_{(1)}^{\top}\mbZ\mbC_{(1)})+tr(\mbC_{(1)}^{\top}\mbZ^{\top}\mbZ\mbC_{(1)}))\\
	& =\frac{1}{2}tr(\mbZ^{\top}\mbZ)+\frac{\lambda_{z}}{2}(tr(\mbC_{(1)}^{\top}\mbC_{(1)})-2tr(\mbC_{(1)}^{\top}\mbZ\mbC_{(1)})+tr(\mbC_{(1)}\mbC_{(1)}^{\top}\mbZ^{\top}\mbZ))
\end{align*}

Taking the derivative of $\mathcal{L_{\mbZ}}$ with respect to $\mbZ$
we get:

\begin{align*}
	\frac{\partial\mathcal{L}_{\mbZ}}{\partial\mbZ} & =\mbZ-\lambda_{z}(\mbC_{(1)}\mbC_{(1)}^{\top})+\lambda_{z}\mbC_{(1)}\mbC_{(1)}^{\top}\mbZ
\end{align*}

Setting $\frac{\partial\mathcal{L}_{\mbZ}}{\partial\mbZ}=\mbzero$,
we find the closed-form solution $\mbZ^{*}$ as in Equation (6).
        \end{proof}
    \subsection{Proof of Proposition 3.3}
\begin{proof}
Rearranging the terms, we get the following objective function:
\begin{align*}
	\mathcal{L}_{\mbC_{(1)}} & \triangleq\frac{\lambda_{z}}{2}\gg\mbC_{(1)}-\mbZ\mbC_{(1)}\gg_{F}^{2}+\frac{\lambda_{e}}{2}\gg\mbM-\mbC_{(1)}\mbP\gg_{F}^{2}\\
	 &\begin{aligned}=\frac{\lambda_{z}}{2}tr(\mbC_{(1)}^{\top}(\mbI-\mbZ)^{\top}(\mbI-\mbZ)\mbC_{(1)})+ \\ \frac{\lambda_{e}}{2}tr(\mbM^{\top}\mbM-2\mbM^{\top}\mbC_{(1)}\mbP+\mbP^{\top}\mbC_{(1)}^{\top}\mbC_{(1)}\mbP)
	\end{aligned}
	\\
	 &\begin{aligned}=\frac{\lambda_{z}}{2}tr(\mbC_{(1)}^{\top}(\mbI-\mbZ)^{\top}(\mbI-\mbZ)\mbC_{(1)})+\frac{\lambda_{e}}{2}tr(\mbM^{\top}\mbM)\\ -\lambda_{e}tr(\mbP\mbM^{\top}\mbC_{(1)})+ \frac{\lambda_{e}}{2}tr(\mbP\mbP^{\top}\mbC_{(1)}^{\top}\mbC_{(1)})
	\end{aligned}
\end{align*}

Taking the derivative of $\mathcal{L}_{\mbC_{(1)}}$ with respect
to $\mbC_{(1)}$ we get:

\[
\frac{\partial\mathcal{L}_{\mbC_{(1)}}}{\partial\mbC_{(1)}}=\lambda_{z}(\mbI-\mbZ)^{\top}(\mbI-\mbZ)\mbC_{(1)}-\lambda_{e}\mbM\mbP^{\top}+\lambda_{e}\mbP\mbP^{\top}\mbC_{(1)}
\]

Here, notice that $\mbP\mbP^{\top}=(\mbU_{3}^{\top}\otimes\mbU_{2}^{\top})(\mbU_{3}\otimes\mbU_{2})=(\mbU_{3}^{\top}\mbU_{3})\otimes(\mbU_{2}^{\top}\mbU_{2})=\mbI$,
because of the properties of the Kronecker product and $\mbU_{2}$
and $\mbU_{3}$ being orthonormal matrices. This helps us simplify the
closed-form solution $\mbC_{(1)}^{*}$ which is found by setting $\frac{\partial\mathcal{L}_{\mbC_{(1)}}}{\partial\mbC_{(1)}}=\mbzero$. 
\end{proof}

    \subsection{Proof of Proposition 3.4}
\begin{proof}
By unfolding the tensor into the 2nd mode, we have
\begin{align*}
 & \underset{\mbU_{2}}{\arg\min}\frac{\lambda_{e}}{2}\|\mathcal{X}-\mathcal{A}-\mathcal{C}\times_{2}\mbU_{2}\times_{3}\mbU_{3}\|_{F}^{2}\\
= & \underset{\mbU_{2}}{\arg\min}\|\mbX_{(2)}-\mbA_{(2)}-\mbU_{2}\mbC_{(2)}(\mbU_{3}\otimes\mbI)^T\|_{F}^{2}\\
= & \underset{\mbU_{2}}{\arg\min}\|\mbB_{1}-\mbU_{2}\mbB_{2}\|_{F}^{2},
\end{align*}
where $\mbB_{1} = \mbX_{(2)}-\mbA_{(2)}$ and $\mbB_2 = \mbC_{(2)}(\mbU_{3}\otimes\mbI)^T$. In this form, this problem has the same structure as the orthogonal Procrustes problem (Sch{\"o}nemann 1966). 

The solution is given by the SVD decomposition on $\mbQ = \mbB_{1} \mbB_2 ^T = (\mbX_{(2)}-\mbA_{(2)})(\mbU_{3}\otimes\mbI)\mbC_{(2)}^{\top}$ as  $\mbQ=\hat{\mbU}\mbD\hat{\mbV}^{\top}$. Finally, the solution is $\mbU_{2}^* = \hat{\mbU}\hat{\mbV}^{\top}$. 
\end{proof}

\end{APPENDICES}

\ACKNOWLEDGMENT{This work is partially funded by the Hong Kong Research Grants Council (RGC) under the grants RGC GRF 16201718 and 16216119.}

\begin{small}
    \bibliography{bibliography.bib}
    \bibliographystyle{informs2014}
\end{small}

\end{document}